\pdfoutput=1
\documentclass[journal,onecolumn,11pt]{IEEEtran}

\usepackage[english]{babel}
\addto\captionsenglish{\def\figurename{Fig.}\def\tablename{TABLE}}
\usepackage{amsmath,amsfonts,amssymb}
\usepackage{amsthm}
\usepackage{algorithm,algorithmic}
\usepackage{graphicx,float}
\usepackage{cite,url}
\usepackage[hidelinks]{hyperref}
\usepackage{cleveref}
\usepackage{thmtools,mdframed,xcolor}
\usepackage{booktabs,multirow,threeparttable}
\usepackage{enumitem}
\usepackage{placeins}
\usepackage{tabularx,array}

\allowdisplaybreaks
\newcommand{\R}{\mathbb{R}}

\theoremstyle{plain}
\newtheorem{theorem}{Theorem}[section]

\newtheorem{proposition}[theorem]{Proposition}
\newtheorem{corollary}[theorem]{Corollary}

\theoremstyle{definition}
\newtheorem{definition}[theorem]{Definition}
\newtheorem{remark}[theorem]{Remark}

\declaretheoremstyle[
    headfont=\bfseries,
    bodyfont=\normalfont,
    mdframed={
        backgroundcolor=blue!5,
        linewidth=1pt,
        roundcorner=5pt,
        innertopmargin=5pt,
        innerbottommargin=5pt
    }
]{defstyle}

\newcommand{\arxivDOI}{10.1109/TSP.2026.3753742}
\newcommand{\arxivIEEEnotice}{\textcopyright~2026 IEEE. Personal use of this material is permitted. Permission from IEEE must be obtained for all other uses, in any current or future media, including reprinting/republishing this material for advertising or promotional purposes, creating new collective works, for resale or redistribution to servers or lists, or reuse of any copyrighted component of this work in other works.}

\title{A Closed-Form Soft Spectral Partition for Adaptive Mode Decomposition}

\author{%
    \IEEEauthorblockN{%
        Hao Jia\IEEEauthorrefmark{1}\IEEEauthorrefmark{2},
        Chunyu Yang\IEEEauthorrefmark{3},
        Fan Feng\IEEEauthorrefmark{1}\IEEEauthorrefmark{2},
        Cesar F. Caiafa\IEEEauthorrefmark{4},
        Feng Duan\IEEEauthorrefmark{2}\IEEEauthorrefmark{5},
        Jordi Sol\'{e}-Casals\IEEEauthorrefmark{6}%
    }

    \IEEEauthorblockA{%
        \IEEEauthorrefmark{1}School of Medicine, Nankai University, Tianjin, China\\
        \IEEEauthorrefmark{2}Tianjin Key Laboratory of Interventional
          Brain-Computer Interface and Intelligent Rehabilitation,
          Nankai University, Tianjin, China\\
        \IEEEauthorrefmark{3}Department of Mechanical Engineering,
          Universitat Polit\`{e}cnica de Catalunya-BarcelonaTech,
          Terrassa, Spain\\
        \IEEEauthorrefmark{4}Instituto Argentino de Radioastronom\'{i}a
          (CCT-CONICET La Plata/CIC-PBA/UNLP), Villa Elisa, Argentina\\
        \IEEEauthorrefmark{5}College of Artificial Intelligence,
          Nankai University, Tianjin, China\\
        \IEEEauthorrefmark{6}Data and Signal Processing Research Group,
          University of Vic-Central University of Catalonia, Vic, Spain%
    }
    \thanks{Accepted for publication in \emph{IEEE Transactions on Signal
      Processing}, DOI: \href{https://doi.org/\arxivDOI}{\arxivDOI}. This is the
      authors' accepted manuscript, not the final published version.
      \arxivIEEEnotice}%
    \thanks{This work was supported in part by the National Key Research and
      Development Program of China under Grant 2025YFE0213700, in part by the Tianjin
      Science and Technology Plan Project under Grants 25ZXWCSY00340 and 25JCQNJC01270,
      in part by the Fundamental Research Funds for the Central Universities under
      Grant 63261198, in part by CONICET (Consejo Nacional de Investigaciones
      Cient\'{\i}ficas y T\'{e}cnicas), Argentina, PIP 11220210100284CO (CFC), and in part
      by the Ministerio de Ciencia e Innovaci\'{o}n (MCIN), Agencia Estatal de
      Investigaci\'{o}n (AEI), Spain, under Grant PID2023-147787OB-I00.}%
    \thanks{Corresponding authors: Feng Duan (email: duanf@nankai.edu.cn) and
      Jordi Sol\'{e}-Casals (email: jordi.sole@uvic.cat).}%
    \thanks{In the published article the proofs and Experiments S1--S5 (Appendices A--I
      here) are supplementary material; in this version they are appended after the main
      text.}%
}

\begin{document}
\maketitle

\begin{abstract}
Density-based Adaptive Mode Decomposition (DAMD) shows that the Variational Mode
Decomposition (VMD) objective, restricted to hard binary partitions, is equivalent to
weighted K-means clustering, enabling automatic mode detection.  The equivalence
concerns detection only; extraction with hard Voronoi masks introduces boundary
artefacts.  This paper develops a soft extraction theory independent of the detector.  First, the hard-partition equivalence is shown to be an algebraic
identity that holds \emph{if and only if} the weights are binary; hard and soft extraction are optima of two
different functionals (linear K-means and regularised quadratic VMD) over the same
simplex.  Second, the regularised soft VMD problem is
proved to have a closed-form global optimum, unique on the support of the
spectrum: the Lorentzian-weighted partition
$s_k^*(\omega)\propto[1+2\alpha(\omega-\omega_k)^2]^{-1}$.  Every fixed point of the VMD
ADMM mode update is proportional to $|\omega-\omega_k|^{-2}$; when the dual
ascent enforces reconstruction, it is the $\alpha\to\infty$ endpoint of this family.
The Lorentzian shape at finite $\alpha$ is therefore a consequence of the regulariser,
and $\alpha$ controls how closely the allocation approaches that fixed point.  Third,
under a Lorentzian spectral prior the partition equals the minimum mean-square error
(MMSE) estimator of each mode in the low-noise limit, tying $\alpha$ to the inverse
squared mode bandwidth.  Since the partition needs only detected centre frequencies, the
extraction rule can be switched without recomputing detection.  Synthetic
experiments with known ground truth test the predictions; four categories of real
signals show how much the two allocations differ.
\end{abstract}


\IEEEpeerreviewmaketitle

\section{Introduction}
\label{sec:intro}

\IEEEPARstart{N}{on-stationary} signals arise in biomedical engineering, fault diagnosis, astrophysics and speech processing
\cite{oppenheim1999discrete}.  Variational Mode Decomposition (VMD)
\cite{dragomiretskiy2014vmd} is the standard mode-decomposition method, owing to its
variational foundation and link to Wiener filtering
\cite{wiener1949extrapolation}.  It needs a preset mode number $K$, and its per-frame
cost grows with $K$ and the iteration count of the Alternating Direction Method
of Multipliers (ADMM) \cite{boyd2011distributed}; Short-Time VMD (STVMD)
\cite{jia_short-time_2025} multiplies this cost by the number of frames~$T$.  Methods automating $K$, such as successive decomposition
\cite{nazari_successive_2020,liu_successive_2022}, information-theoretic criteria
\cite{lian_adaptive_2018,xu_adaptive_2023}, metaheuristics
\cite{wang_adaptive_2022,yan_adaptive_2024} and density-based detection
\cite{gu_time-frequency_2025}, all retain the iterative ADMM.
Density-based Adaptive Mode Decomposition (DAMD) \cite{jia2025damd}
addressed this bottleneck by proving that, under hard (binary) allocation,
VMD optimisation is equivalent to weighted K-means clustering on the power spectrum,
so that a single clustering pass (e.g., mean shift) replaces ADMM and also determines $K$.

This is only the \emph{forward direction} (binary weights, detection by clustering); the \emph{converse}, whether the equivalence requires hard partition, was
left open.  Yet VMD's ADMM fixed points satisfy
$\hat U_k\propto d_k^{-2}$ (Section~\ref{sec:background}), neither the Voronoi mask nor
the normalised Lorentzian \eqref{eq:wiener_weights}.  Soft spectral allocation has
thus lacked a variational or statistical justification within DAMD, while hard
extraction creates step discontinuities at mode boundaries, hence Gibbs-like artefacts
and energy misattribution when modes overlap.
Of \emph{detection} (how many modes, and where) and \emph{extraction} (how energy
is allocated among them), DAMD
addressed detection via the binary-restricted clustering equivalence, but its
extraction inherits the hard Voronoi mask, the minimiser of that surrogate and not
necessarily a good choice for extraction.

Our soft partition needs only the detected centres, so hard and soft extraction are
interchangeable without recomputing detection.  The three main results are:

\begin{enumerate}
\item \emph{Hard partition is necessary for the equivalence}
(Theorem~\ref{thm:algebraic}): The VMD and K-means objectives are
pointwise equal if and only if the weights are binary, the only ones satisfying
$s_k^2 = s_k$; any soft allocation on a positive-energy region strictly lowers the
VMD objective below the K-means value.

\item \emph{The optimal soft partition}
(Theorem~\ref{thm:soft_opt}): The regularised soft VMD problem has a
closed-form global optimum, unique on the support of the spectrum:
the Lorentzian-weighted partition, computed in a single pass.
The VMD ADMM fixed point ($\propto|\omega-\omega_k|^{-2}$) with reconstruction
enforced is the $\alpha\to\infty$ endpoint of this family; finite $\alpha$ adds a
prior bandwidth, which yields the Lorentzian shape.

\item \emph{The statistical meaning of the Lorentzian weights}
(Theorem~\ref{thm:mmse}): Under a Lorentzian spectral prior, the
soft partition of Theorem~\ref{thm:soft_opt} equals the MMSE estimator of each
mode in the low-noise limit, tying $\alpha$ to the inverse squared mode bandwidth.
\end{enumerate}

Supporting results include the soft M-step (Proposition~\ref{prop:mstep}), the
monotone decrease of the regularised objective under the soft EM iteration
(Proposition~\ref{prop:conv}), a closed-form FWHM-based
bandwidth estimator (Corollary~\ref{cor:fwhm}), and the limits of the Lorentzian family
($\alpha\to0$: uniform; $\alpha\to\infty$: inverse-square, not the hard Voronoi mask,
which optimises a different, linear functional).  Sections~\ref{sec:background}--\ref{sec:algorithm}
give theory and algorithm; Sections~\ref{sec:experiments}--\ref{sec:conclusion},
experiments and discussion.

\section{Background}
\label{sec:background}

\subsection{Variational Mode Decomposition}
\label{subsec:vmd}

VMD \cite{dragomiretskiy2014vmd} decomposes a signal $x(t)$ into $K$
band-limited modes $\{u_k(t)\}_{k=1}^K$ by solving
\begin{align}
  &\min_{\{U_k\},\{\omega_k\}}
  \sum_{k=1}^K \int_\R |\omega-\omega_k|^2 |U_k(\omega)|^2 \,d\omega
  \notag\\
  &\text{s.t.} \;\sum_{k=1}^K U_k(\omega) = X(\omega).
\label{eq:vmd_obj}
\end{align}
where $X(\omega)$ and $U_k(\omega)$ are the spectra of $x(t)$ and $u_k(t)$
(analytic signals).

\textbf{Frequency convention.}
Following~\cite{dragomiretskiy2014vmd}, Sections~\ref{sec:background}--\ref{sec:convergence}
write $\omega$ for frequency; the analytic-signal spectra $X$, $U_k$ vanish for $\omega<0$, so integrals over~$\mathbb{R}$
reduce to~$[0,\infty)$, as in~\eqref{eq:admm_freq}.
In the implementation and all experiments, computations use the \emph{normalised}
frequency $\omega=f/f_s\in[0,\,0.5]$, with
$F=\lfloor n_{\mathrm{fft}}/2\rfloor+1$ one-sided bins, STFT bin $m$ lying at
$\omega=m/n_\mathrm{fft}$; physical frequencies are reported in hertz where
convenient.
The Lorentzian denominator is $1+2\alpha(\omega-\omega_k)^2$ throughout
(the VMD reference code uses $1+\alpha(\omega-\omega_k)^2$,
so $\alpha_{\mathrm{impl}}=2\alpha$; with frequency in hertz, or as $2\pi\omega$ in
rad/sample, $\alpha$ becomes $\alpha/f_s^2$, or $\alpha/(2\pi)^2$).
The half-bandwidth is $\beta_k=1/\sqrt{2\alpha_k}$ in units of~$\omega$; the default
$\alpha=2000$ gives $\beta\approx0.016$ normalised, i.e.\ $15.8$\,Hz at $f_s=1000$\,Hz
and $3.95$\,Hz at $f_s=250$\,Hz.

ADMM \cite{boyd2011distributed} iterates, with penalty (bandwidth) parameter $\alpha>0$
and multiplier $\hat{\lambda}^{(n)}(\omega)$ enforcing reconstruction,
\begin{equation}
  \hat{U}_k^{(n+1)}(\omega)
  = \frac{\hat{X}(\omega) - \sum_{i\neq k}\hat{U}_i^{(n)}(\omega)
          + \hat{\lambda}^{(n)}(\omega)/2}
         {1 + 2\alpha(\omega-\omega_k^{(n)})^2},
\label{eq:admm_mode}
\end{equation}
\begin{equation}
  \omega_k^{(n+1)}
  = \frac{\int_0^\infty \omega|\hat{U}_k^{(n+1)}(\omega)|^2\,d\omega}
         {\int_0^\infty |\hat{U}_k^{(n+1)}(\omega)|^2\,d\omega},
\label{eq:admm_freq}
\end{equation}
until convergence.

At a fixed point of \eqref{eq:admm_mode}, with primal residual
$R(\omega)=\hat X(\omega)-\sum_j \hat U_j(\omega)$, its
numerator is $R+\hat U_k+\hat\lambda/2$, so
\begin{equation}
  2\alpha\,d_k^2(\omega)\,\hat U_k(\omega)
  \;=\; R(\omega)+\tfrac12\hat\lambda(\omega)
  \;=:\;Q(\omega),
\label{eq:admm_fp_identity}
\end{equation}
where $d_k\triangleq|\omega-\omega_k|$.  The right-hand side is independent of $k$, so
every fixed point satisfies
$\hat U_k\propto d_k^{-2}$, with no assumption on $\hat\lambda$ or on mode overlap.
When the multiplier enforces reconstruction ($R=0$), summing
\eqref{eq:admm_fp_identity} over $k$ fixes the normalisation, for
$\omega\notin\{\omega_1,\dots,\omega_K\}$:
\begin{equation}
  \hat{U}_k^*(\omega) = \frac{d_k^{-2}(\omega)}{\sum_j d_j^{-2}(\omega)}\cdot X(\omega).
\label{eq:vmd_wiener}
\end{equation}
At a centre the ratio is undefined and \eqref{eq:admm_fp_identity} applies
directly: $\omega=\omega_k$ forces $Q=0$, hence $\hat U_j=0$ for every
$\omega_j\neq\omega_k$ and, by reconstruction, $\hat U_k=X$.  Where $m\geq2$ centres coincide only their sum is fixed, and
the equal sharing of Corollary~\ref{cor:hard_limit} follows from the Lorentzian
limit rather than from the fixed-point condition.
Equation~\eqref{eq:vmd_wiener} is the $\alpha\to\infty$ endpoint of the Lorentzian family of
Section~\ref{sec:soft_opt} (Corollary~\ref{cor:hard_limit}; see
Remark~\ref{rem:admm_fp}), and \emph{not} the normalised Lorentzian
\begin{equation}
  w_k(\omega)
  \triangleq
  \frac{[1+2\alpha(\omega-\omega_k)^2]^{-1}}
       {\sum_{j=1}^K [1+2\alpha(\omega-\omega_j)^2]^{-1}},
\label{eq:wiener_weights}
\end{equation}
which is the renormalised first Jacobi update of \eqref{eq:admm_mode} from
$\hat U_i\equiv0$.

\textbf{Scope of \eqref{eq:admm_fp_identity}--\eqref{eq:vmd_wiener}.}
(i)~VMD updates modes sequentially (Gauss--Seidel), but at a fixed point refreshed and
previous values coincide, so \eqref{eq:admm_fp_identity} holds for sequential and
parallel (Jacobi) updates alike.  (ii)~The multiplier follows dual ascent,
$\hat\lambda^{(n+1)}=\hat\lambda^{(n)}+\tau\bigl(\hat X-\sum_k\hat U_k^{(n+1)}\bigr)$
\cite{dragomiretskiy2014vmd}; for $\tau>0$ a fixed point requires $R=0$, giving
\eqref{eq:vmd_wiener} with a multiplier
$\hat\lambda=4\alpha\hat X/\sum_jd_j^{-2}$ that is in general non-zero.  (iii)~With
$\tau=0$ the multiplier stays at zero, $Q=R$, and summing
\eqref{eq:admm_fp_identity} gives
$\hat U_k=\bigl[d_k^{-2}/\sum_jd_j^{-2}\bigr]\cdot\bigl[A/(1+A)\bigr]\hat X$ with
$A=\sum_j(2\alpha d_j^2)^{-1}$: the same inverse-square shape times a common shrinkage
$A/(1+A)<1$, which does not reconstruct $\hat X$.  Experiment~A
checks both cases against the solver.
\textbf{Terminology.}
We call~\eqref{eq:wiener_weights} the \emph{Lorentzian mask} (or
\emph{normalised Lorentzian weights}); \emph{Wiener/MMSE} is reserved for the estimator
$S_k(\omega)/S_X(\omega)$ of the stochastic model of Section~\ref{sec:mmse}, which
coincides with it only in the low-noise limit (Theorem~\ref{thm:mmse}).
Note $\sum_k w_k(\omega)=1$ and $w_k(\omega)\in(0,1)$ for finite $\alpha$.

STVMD \cite{jia_short-time_2025} applies VMD independently to each frame
$t \in \{1,\ldots,T\}$, giving centres $\{\omega_{k,t}\}$; it is the cost baseline of
Section~\ref{sec:algorithm}.

\subsection{DAMD and the Equivalence Theorem}
\label{subsec:damd}

\subsubsection{Parameterisation}
Write mode extraction as $U_k(\omega) = s_k(\omega)X(\omega)$ with allocation weight
$s_k(\omega)\geq 0$; reconstruction requires $\sum_k s_k(\omega)=1$.  DAMD uses the
hard (binary) allocation
\begin{equation}
  s_k(\omega) = \mathbf{1}_{B_k}(\omega)
  \triangleq
  \begin{cases}1 & \text{if } k=\arg\min_j|\omega-\omega_j|,\\
               0 & \text{otherwise,}\end{cases}
\label{eq:hard_alloc}
\end{equation}
where $\{B_k\}$ are the Voronoi cells of $\{\omega_k\}$, with ties shared equally:
if $m$ modes are equidistant from~$\omega$, each receives $s_k(\omega)=1/m$, preserving
$\sum_k s_k=1$.  Sharing is a convention of the discrete
implementation, outside the binary restriction: the tie set has measure zero for
distinct centres, and off it \eqref{eq:hard_alloc} is binary as
Proposition~\ref{thm:damd_equiv} and Theorem~\ref{thm:algebraic} assume.

\subsubsection{VMD–Clustering Equivalence}
Under \eqref{eq:hard_alloc}, \eqref{eq:vmd_obj} reduces to weighted K-means:

\begin{proposition}[VMD--Clustering Equivalence \cite{jia2025damd}]
\label{thm:damd_equiv}
Restricted to binary allocations $U_k=s_kX$ with $s_k(\omega)\in\{0,1\}$ and
$\sum_ks_k=1$, the VMD optimisation \eqref{eq:vmd_obj} is equivalent to
\begin{equation}
  \min_{\{\omega_k\}_{k=1}^K}
  \int_\R \min_k |\omega-\omega_k|^2 \,|X(\omega)|^2\,d\omega,
\label{eq:kmeans_obj}
\end{equation}
with optimal modes given by the hard partition
$U_k^*(\omega)=X(\omega)\cdot\mathbf{1}_{B_k}(\omega)$.
\end{proposition}

\begin{remark}[Scope of Proposition~\ref{thm:damd_equiv}]
The binary restriction is essential: without it the fixed-centre problem
$\min_{s\in\Delta^K}\sum_k d_k^2s_k^2P$ has the inverse-square allocation as its
stationary point (Remark~\ref{rem:estep_obj}), not the Voronoi mask.
Proposition~\ref{thm:damd_equiv} characterises the binary-restricted surrogate that
motivates DAMD's \emph{detection}; no optimality claim transfers to the
unrestricted problem.
\end{remark}

\subsubsection{Automatic Mode Detection via Density-Based Clustering}
Accordingly, DAMD replaces ADMM with a single density-based clustering pass
\cite{comaniciu2002mean,cheng1995mean,fukunaga1975estimation} on
$|X(\omega)|^2$, counting density peaks to obtain $K^*$.
The pass is motivated by \eqref{eq:kmeans_obj} but does not solve it: it locates
maxima of a kernel density estimate of the power spectrum, and its centres are not
claimed to minimise \eqref{eq:kmeans_obj}.  We thus keep apart the unrestricted
fixed-centre VMD problem (inverse-square stationary allocation), the binary-restricted
surrogate of Proposition~\ref{thm:damd_equiv}, and the clustering algorithm supplying
the centres.  The per-frame complexity is
$O(S\cdot N_\mathrm{M}\cdot F)$, with $S\leq F$ mean-shift seeds and $N_\mathrm{M}$
substantially smaller than ADMM's iteration count~\cite{jia2025damd}.

\subsection{Soft Spectral Allocation and Bayesian Estimation}
\label{subsec:soft_related}

Soft allocation is standard in speech-separation ratio
masks~\cite{srinivasan2006binary}, a graded relaxation of the ideal binary
mask~\cite{wang2005ideal}, and
in Wiener filtering~\cite{wiener1949extrapolation}, the minimum mean-square error (MMSE)
estimator in additive noise.  VMD's link to Wiener filtering was noted
in~\cite{dragomiretskiy2014vmd}.  Bayesian (posterior-weighted)
spectral estimation uses parametric priors~\cite{kay1993fundamentals}; the Lorentzian
(Cauchy) shape arises in damped resonances~\cite{papoulis2002probability}.
As in mixture models fitted by Expectation-Maximisation
(EM)~\cite{dempster1977maximum}, our soft partition admits an EM interpretation (Section~\ref{sec:convergence}).  No prior work has derived
a soft spectral partition as the optimum of a variational problem (one
whose solution \emph{equals} the Lorentzian weights), connected it to MMSE estimation,
or related it to DAMD's hard partition.

\subsection{The Hard--Soft Tension}
\label{subsec:tension}

A hard partition is globally optimal within the \emph{hard-partition/clustering
formulation} (Proposition~\ref{thm:damd_equiv}), whereas the ADMM
update~\eqref{eq:admm_mode} divides by $1+2\alpha d_k^2$ and its fixed points are the
soft allocation~\eqref{eq:vmd_wiener} once the multiplier enforces
reconstruction.  Two questions follow: why the hard partition is
equivalent to the K-means objective, and whether the Lorentzian weighting
\eqref{eq:wiener_weights} is optimal for \emph{some} well-posed problem and how it
relates to the hard-partition result.

\section{Algebraic Foundation of the VMD--Clustering Equivalence}
\label{sec:algebraic}

\subsection{Parametric Form of the Two Objectives}

Substituting $U_k = s_k X$ into the VMD objective \eqref{eq:vmd_obj}, and
writing the K-means cost \eqref{eq:kmeans_obj} with an explicit allocation variable in
place of its pointwise minimum, yields
\begin{align}
  J_\mathrm{VMD}   &= \sum_{k=1}^K \int_\R
    |\omega-\omega_k|^2 s_k^2(\omega) |X(\omega)|^2\,d\omega,
  \label{eq:Jvmd}\\
  J_\mathrm{Kmeans}&= \sum_{k=1}^K \int_\R
    |\omega-\omega_k|^2 s_k(\omega) |X(\omega)|^2\,d\omega.
  \label{eq:Jkm}
\end{align}
Only the exponent of $s_k$ differs.
Equation~\eqref{eq:Jkm} is the assignment-variable form of
\eqref{eq:kmeans_obj}, not a substitution into it: as in
Proposition~\ref{thm:damd_equiv}, at fixed centres its minimum over binary $s$ is the
integrand $\min_k|\omega-\omega_k|^2$ of \eqref{eq:kmeans_obj}, and over the simplex
it is the linear relaxation.  Below, \eqref{eq:Jvmd} and \eqref{eq:Jkm} are compared as functionals of $s$ at fixed
centres; \eqref{eq:Jkm} is never identified with \eqref{eq:kmeans_obj} at a
non-binary $s$.

\subsection{Main Result}

\begin{theorem}[Algebraic Identity]
\label{thm:algebraic}
Fix a centre configuration $\{\omega_k\}$ and allocation functions
$s_k:[0,\infty)\to[0,1]$ (the analytic-spectrum domain) with $\sum_k s_k\equiv1$
that do not depend on the signal.
The integrand-level identity $|\omega - \omega_k|^2 s_k^2(\omega) \equiv |\omega - \omega_k|^2 s_k(\omega)$ holds for all $k$ and almost all $\omega$ \emph{if and only if} $s_k(\omega) \in \{0, 1\}$ for all $k$ and almost all $\omega$.  Equivalently, the integral identity $J_\mathrm{VMD} = J_\mathrm{Kmeans}$ holds for every signal $X$ if and only if $s_k(\omega) \in \{0, 1\}$ for all $k$ and almost all $\omega$.
The allocation being fixed in advance is part of the statement, and the
quantifier over $X$ is necessary in the converse: since
$J_\mathrm{VMD}-J_\mathrm{Kmeans}=\sum_k\int d_k^2(s_k^2-s_k)P\,d\omega$ with every
integrand non-positive, equality for one signal only forces
$s_k(\omega)\in\{0,1\}$ where $P(\omega)>0$ and $d_k(\omega)>0$; an allocation
non-binary only where the signal carries no energy gives equality for that
signal.
\end{theorem}

\emph{Proof:} see Appendix~\ref{app:proofs}.

\begin{corollary}[Strict Inequality for Soft Allocations]
\label{cor:strict}
If there exists a set of positive measure on which $P(\omega)>0$,
$|\omega-\omega_k|>0$, and $s_k(\omega)\in(0,1)$ for at least one
$k$, then $J_\mathrm{VMD} < J_\mathrm{Kmeans}$ strictly.
\end{corollary}

\emph{Proof:} see Appendix~\ref{app:proofs}.

\subsection{Geometric Interpretation}

The feasible set $\Delta^K = \{s\in\R^K : s_k\geq 0,\,\sum_k s_k=1\}$
is the $(K-1)$-simplex, and the hard allocations \eqref{eq:hard_alloc}
are its $K$ vertices, the only points with $s_k^2=s_k$.  By
Theorem~\ref{thm:algebraic} and Corollary~\ref{cor:strict}, the
equivalence applies to the \emph{vertices} of the feasible set, and any
interior point on a positive-energy region strictly lowers $J_\mathrm{VMD}$ below
$J_\mathrm{Kmeans}$; for $K\geq3$ the simplex boundary contains many non-binary
points, where Corollary~\ref{cor:strict} already applies, so ``boundary'' would
overstate the equivalence.
Other points need the framework developed next.

\section{Optimal Soft Spectral Partition}
\label{sec:soft_opt}

\subsection{Motivation: Failure Modes of Hard Partition}
\label{subsec:hard_failures}

Hard partition can be suboptimal in two common scenarios.
\textbf{Overlapping modes.}  When two centres are closer than the
clustering bandwidth, the Voronoi boundary bisects shared energy, assigning
bins by proximity alone, and the step mask causes Gibbs-like ringing.

\textbf{Time-varying signals.}  As centres evolve, the hard boundary
may jump between frames $t$ and $t+1$, creating discontinuities absent
from the physical oscillation.
As labels are not associated across frames (Section~\ref{sec:conclusion}),
this is motivation only; what is guaranteed is that, for a fixed number of centres,
each soft weight varies continuously with the centres, whereas a hard weight jumps
when a boundary crosses its bin.

\subsection{The Regularised Soft VMD Problem}

\begin{definition}[Regularised Soft VMD]
\label{def:soft_vmd}
Given mode centres $\{\omega_k\}_{k=1}^K$ and penalty parameter
$\alpha>0$, the regularised soft VMD problem is
\begin{equation}
  \tilde J_\alpha(\{s_k\};\{\omega_k\})
  \;\triangleq\;
  \sum_{k=1}^K \int_\R
    \underbrace{\Bigl(|\omega-\omega_k|^2+\tfrac{1}{2\alpha}\Bigr)}_
    {\textstyle c_k(\omega)\,>\,0}
    s_k^2(\omega)\,P(\omega)\,d\omega
\label{eq:soft_vmd}
\end{equation}
to be minimised over $\{s_k\}$ subject to $\sum_k s_k(\omega)=1$ and
$s_k(\omega)\geq 0$ for all $\omega$, where $P(\omega)=|X(\omega)|^2$.
\end{definition}

The term $1/(2\alpha)$ is one choice of offset, not the minimal one: any
positive constant makes $c_k>0$, removing the singularity at $\omega=\omega_k$ and
making the solution well-defined even for coincident centres.

\begin{remark}[Choice of regulariser]
\label{rem:regulariser}
The regulariser $1/(2\alpha)$ with the $s_k^2$ exponent is one of several natural choices; an entropy regulariser $-\sum_k s_k \log s_k$ would yield a softmax allocation $s_k^* \propto \exp(-\kappa |\omega - \omega_k|^2)$.  This combination is the one whose minimiser is the normalised Lorentzian
mask \eqref{eq:wiener_weights}; that coincidence is a consequence of the choice rather
than evidence for it, and the independent motivation is the Bayesian one of
Section~\ref{sec:mmse}.
Replacing $s_k^2$ by $s_k^p$, $p>1$, gives $s_k^*\propto c_k^{-1/(p-1)}$; an
offset $\epsilon>0$ in place of $1/(2\alpha)$ gives a Lorentzian-type mask of
half-width $\sqrt\epsilon$, and $\epsilon\to0$ the inverse-square allocation
\eqref{eq:unreg_min}.  Section~\ref{sec:mmse} selects $p=2$ with $\epsilon$ the
squared prior half-width, giving the low-noise MMSE weights
(Theorem~\ref{thm:mmse}).
\end{remark}

\subsection{Closed-Form Optimal Solution}

\begin{theorem}[Closed-form optimum of regularised soft VMD]
\label{thm:soft_opt}
On the support $\{\omega:P(\omega)>0\}$ the
problem~\eqref{eq:soft_vmd} is strictly convex in $\{s_k(\omega)\}$
for each $\omega$ independently, with unique global minimiser
\begin{equation}
  s_k^*(\omega)
  = \frac{[1+2\alpha(\omega-\omega_k)^2]^{-1}}
         {\displaystyle\sum_{j=1}^K[1+2\alpha(\omega-\omega_j)^2]^{-1}}
\label{eq:soft_opt}
\end{equation}
for $k=1,\ldots,K$.  Where $P(\omega)=0$ the per-frequency objective
vanishes for any feasible $s_k$; we adopt~\eqref{eq:soft_opt} there by
convention, extending $\{s_k^*\}$ to all frequencies without changing
$\tilde J_\alpha$.
\end{theorem}

\emph{Proof:} see Appendix~\ref{app:proofs}.

\begin{remark}
If $\omega_i=\omega_j$, then $c_i=c_j>0$ and $s_i^*=s_j^*$ (equal
sharing); uniqueness is preserved.
\end{remark}

\begin{remark}[Coincidence with the normalised Lorentzian mask, by construction]
\label{cor:vmd_exact}
The optimal soft partition \eqref{eq:soft_opt} is identical to the normalised
Lorentzian mask \eqref{eq:wiener_weights}:
\begin{equation}
  s_k^*(\omega;\alpha) \;\equiv\; w_k(\omega;\alpha) \qquad
  \forall\,\omega,\,\forall\,\alpha>0.
\end{equation}
The identity is definitional: the regulariser of Definition~\ref{def:soft_vmd} was chosen so that the inverse
cost reproduces the denominator of \eqref{eq:wiener_weights}
(Remark~\ref{rem:regulariser}).  It shows one object as a mask and as a minimiser,
and makes no claim about the ADMM output.
\end{remark}

\begin{remark}[What the identity does and does not assert]
Remark~\ref{cor:vmd_exact} does not state that \eqref{eq:soft_opt}
reproduces the VMD ADMM output: by \eqref{eq:admm_fp_identity} every
fixed point of \eqref{eq:admm_mode} is proportional to $d_k^{-2}$, so the family meets
it only at its $\alpha\to\infty$ endpoint (Fig.~\ref{fig:exp_a}).
It records that a Wiener-analogy mask minimises an explicit objective
(Theorem~\ref{thm:soft_opt}); since that objective was chosen to produce it
(Remark~\ref{rem:regulariser}), the independent justification is Bayesian
(Theorem~\ref{thm:mmse}).  Neither depends on ADMM.
\end{remark}

\begin{corollary}[Weighted Generalisation]
\label{cor:weighted}
Let each mode carry a prior amplitude weight $\sigma_k^2 > 0$ and
a mode-specific bandwidth $\alpha_k > 0$, and write
$\beta_k^2\triangleq1/(2\alpha_k)$.  The weighted regularised
soft VMD problem
\begin{equation}
  \min_{\{s_k\}} \sum_{k=1}^K \int_\R
    \frac{|\omega-\omega_k|^2+\beta_k^2}{\sigma_k^2\,\beta_k^2}\,
    s_k^2(\omega)\,P(\omega)\,d\omega
\label{eq:weighted_soft_vmd}
\end{equation}
subject to $\sum_k s_k(\omega)=1$, $s_k(\omega)\geq 0$, has, on the support
$\{\omega:P(\omega)>0\}$, the unique global optimal solution
\begin{equation}
  s_k^*(\omega)
  = \frac{\sigma_k^2\,[1+2\alpha_k(\omega-\omega_k)^2]^{-1}}
         {\displaystyle\sum_{j=1}^K
          \sigma_j^2\,[1+2\alpha_j(\omega-\omega_j)^2]^{-1}},
\label{eq:weighted_opt}
\end{equation}
extended by convention to $P(\omega)=0$, as in
Theorem~\ref{thm:soft_opt}.
\end{corollary}

\emph{Proof:} see Appendix~\ref{app:proofs}.

\begin{remark}[Correction to the weighting]
\label{rem:weighted_fix}
Dividing the cost in \eqref{eq:weighted_soft_vmd} by $\sigma_k^2$ alone gives
$\tilde c_k=(d_k^2+\beta_k^2)/\sigma_k^2=(1+2\alpha_kd_k^2)/(2\alpha_k\sigma_k^2)$,
so $s_k^*\propto\tilde c_k^{-1}$ carries an extra factor $\alpha_k$ absent from
\eqref{eq:weighted_opt}; $2\alpha_k$ cancels only for equal bandwidths.  For
$\omega_1=0$, $\omega_2=1$, $\omega=0.5$, $\sigma_1^2=\sigma_2^2=1$, $\alpha_1=1$ and
$\alpha_2=10$, \eqref{eq:weighted_opt} gives $s_1=0.800$ whereas the minimiser with
the $\sigma_k^2$ normalisation is $s_1=0.286$.  Dividing by $\sigma_k^2\beta_k^2$, as
above, makes \eqref{eq:weighted_opt} exact; \eqref{eq:soft_opt} is unaffected, since
there $\alpha_k\equiv\alpha$ is a common factor.
\end{remark}

\begin{corollary}[Limits of the Lorentzian family]
\label{cor:hard_limit}
For any $\omega\notin\{\omega_1,\ldots,\omega_K\}$,
\begin{equation}
  \lim_{\alpha\to\infty} s_k^*(\omega;\alpha)
  \;=\; \frac{|\omega-\omega_k|^{-2}}
              {\sum_{j=1}^K |\omega-\omega_j|^{-2}},
\label{eq:invsq_limit}
\end{equation}
the \emph{inverse-square soft allocation}.  At $\omega=\omega_k$
(centre of mode $k$), and provided the centres are distinct
($\omega_j\neq\omega_k$ for $j\neq k$), $s_k^*(\omega_k)\to 1$ and
$s_j^*(\omega_k)\to 0$ for $j\neq k$.  In the opposite limit,
$\lim_{\alpha\to 0}s_k^*(\omega;\alpha)=1/K$ (uniform allocation).
If $m$ centres coincide at $\omega$, their numerators all equal one and the
limit is $1/m$ for each, consistent with the equal sharing noted after
Theorem~\ref{thm:soft_opt}.
\end{corollary}

\emph{Proof:} see Appendix~\ref{app:proofs}.

\begin{remark}[The $\alpha\to\infty$ limit is not the Voronoi indicator]
\label{rem:not_voronoi}
The limit~\eqref{eq:invsq_limit} differs qualitatively from
the Voronoi indicator $\mathbf{1}_{B_k}$ of DAMD's hard extraction.  With
$\omega_1=0$, $\omega_2=2$, $\omega=0.9$, the Voronoi mask gives
$\mathbf{1}_{B_1}=1$, whereas the limit gives
$0.81^{-1}/(0.81^{-1}+1.21^{-1})\approx0.60$, approached
once $2\alpha d_j^2\gg1$ for every $j$ (the $\alpha$-dependence cancels in the
asymptotic expansion, leaving an $O(1/\alpha)$ residual; see proof).
The hard mask, optimum of the \emph{linear} K-means functional
(Proposition~\ref{thm:damd_equiv}), is thus \emph{not} in the family, which optimises
the \emph{quadratic} functional~\eqref{eq:soft_vmd}; by Theorem~\ref{thm:algebraic}
the K-means and VMD objectives agree only at the vertices.
\end{remark}

Across $\alpha$ the Lorentzian family thus spans
\begin{equation}
\underbrace{s_k^*\to\tfrac{1}{K}}_{\alpha\to 0,\;\text{uniform}}
\;\longrightarrow\;
\underbrace{s_k^*=w_k}_{\text{finite }\alpha,\;\text{Lorentzian mask}}
\;\longrightarrow\;
\underbrace{s_k^*\to d_k^{-2}/\textstyle\sum_j d_j^{-2}}
           _{\alpha\to\infty,\;\text{inverse-square soft}}
\label{eq:spectrum}
\end{equation}
with the hard Voronoi mask \emph{outside} it (Remark~\ref{rem:not_voronoi}).

\section{Bayesian Interpretation: MMSE Optimality}
\label{sec:mmse}

\subsection{Lorentzian Mode Model}

The MMSE result below assumes a Lorentzian spectral prior for each mode, an idealisation that real modes (SSVEP harmonics, fault transients, formants) violate.  It serves two purposes: (i)~it is a family for which the MMSE estimator $S_k(\omega)/S_X(\omega) \cdot X(\omega)$ takes \emph{exactly} the form of $s_k^*$ (Corollary~\ref{cor:weighted} with $\alpha_k = 1/(2\beta_k^2)$, in the low-noise limit); (ii)~it reads the bandwidth parameter as a prior mode width.  For other shapes the departure from MMSE is not small.  Experiment~C$'$ quantifies this for Gaussian,
raised-cosine and rectangular spectra: at $40$\,dB the nMSE is $2.7$ to $9.5$ times the
oracle's, yet at exact centres the mask still beats hard extraction, an advantage lost at
high SNR once the centres are also in error, or for line components far narrower than
the prior (Experiment~S4).

\begin{definition}[Lorentzian Mode Model]
\label{def:lorentz_model}
The observed spectrum is
\begin{equation}
  X(\omega) = \sum_{k=1}^K U_k^\mathrm{true}(\omega) + N(\omega),
\label{eq:obs_model}
\end{equation}
where the modes $\{U_k^\mathrm{true}\}$ are mutually independent
zero-mean complex Gaussian processes with Lorentzian power spectral
densities
\begin{equation}
  S_k(\omega) = \frac{\sigma_k^2}{1+[(\omega-\omega_k)/\beta_k]^2},
\label{eq:lorentz_psd}
\end{equation}
where $\sigma_k^2$ is the \emph{peak} spectral density of mode $k$, not its
integrated power, and $\beta_k>0$ is its Lorentzian
half-bandwidth.  Its integrated power over the band $[0,\tfrac12]$ used here is
$\sigma_k^2\beta_k[\arctan(\tfrac{1/2-\omega_k}{\beta_k})+\arctan\tfrac{\omega_k}{\beta_k}]$,
close to the whole-line value $\pi\sigma_k^2\beta_k$ only for $\beta_k$ small against the
distance to both edges; on the grid, it is the sum over bins.  The noise term $N(\omega)$ is zero-mean complex Gaussian, independent of the
modes and across bins, with flat spectral density $\sigma_N^2$.
As in the experiments, $\omega$ ranges over the $F$
transform bins, the $\{U_k^\mathrm{true}(\omega)\}$ are independent across
modes and bins, and $\{\omega_k\}$, $\{\sigma_k^2\}$, $\{\beta_k\}$,
$\sigma_N^2$ are known.  The penalty parameter
for mode~$k$ is $\alpha_k = 1/(2\beta_k^2)$; the common case
$\alpha_k = \alpha$ suits modes of similar width.
\end{definition}

\subsection{MMSE Estimation}

\begin{theorem}[MMSE Equivalence]
\label{thm:mmse}
Under Definition~\ref{def:lorentz_model}, the MMSE estimator of mode $k$ given $X(\omega)$, with all parameters known, is
\begin{equation}
  \hat{U}_k^\mathrm{MMSE}(\omega) = \frac{S_k(\omega)}{S_X(\omega)}\, X(\omega), \qquad S_X(\omega) = \sum_{j=1}^K S_j(\omega) + \sigma_N^2.
\label{eq:mmse_est}
\end{equation}
In the low-noise limit $\sigma_N^2 \to 0$, $S_X(\omega) \to \sum_j S_j(\omega)$ and
\begin{equation}
  \hat{U}_k^\mathrm{MMSE}(\omega) \;\to\; s_k^*(\omega) \cdot X(\omega),
\label{eq:mmse_limit}
\end{equation}
where $s_k^*$ is the weighted optimal soft partition~\eqref{eq:weighted_opt} with $\alpha_k = 1/(2\beta_k^2)$.  When $\sigma_k^2 = \sigma^2$ and $\alpha_k=\alpha$ for all $k$, this reduces to the equal-weight partition~\eqref{eq:soft_opt}.
Both conditions are required: \eqref{eq:weighted_opt} contains $\sigma_k^2$ and
$\alpha_k$, and equal amplitudes remove only the former.
\end{theorem}

\emph{Proof:} see Appendix~\ref{app:proofs}.

\begin{remark}
Corollary~\ref{cor:weighted}, with the objective corrected as in Remark~\ref{rem:weighted_fix}, is
essential: its solution matches the MMSE Wiener weights \emph{exactly}
without equal prior variances or bandwidths.  Peak levels $\sigma_k^2$ and
bandwidths $\alpha_k$ can be estimated from the peak of each detected band and
its FWHM (Corollary~\ref{cor:fwhm}); a band energy must first be divided by that
integral at unit peak, $\pi\beta_k$ only in the narrow-band limit above.
Theorem~\ref{thm:soft_opt}, Remark~\ref{cor:vmd_exact},
Corollary~\ref{cor:hard_limit}, Remark~\ref{prop:rate}, the Type-II
gradient~\eqref{eq:mle_grad} (also for a common $\sigma^2$) and the ADMM analysis of
Section~\ref{sec:background} assume a common $\alpha$.  Propositions~\ref{prop:mstep}
and~\ref{prop:conv} carry over to per-mode $\alpha_k$ and $\sigma_k^2$ with
Corollary~\ref{cor:weighted} in place of Theorem~\ref{thm:soft_opt}, since the weighted
cost scales each centroid only by the constant $1/(\sigma_k^2\beta_k^2)$.
\end{remark}

\begin{remark}[Finite-noise regime]
\label{rem:finite_noise}
For $\sigma_N^2 > 0$, the MMSE estimator~\eqref{eq:mmse_est} differs from $s_k^* \cdot X$ by the shrinkage factor
\begin{equation}
\frac{S_k(\omega)/S_X(\omega)}{s_k^*(\omega)} = \frac{\sum_j S_j(\omega)}{\sum_j S_j(\omega) + \sigma_N^2} \;\leq\; 1,
\label{eq:shrinkage}
\end{equation}
which approaches $1$ in the mode bands and $0$ between them, so unless $\sigma_N^2 \to 0$, $s_k^*$ is tight at the mode frequencies and biased between modes.  The gap is not negligible: in Experiment~C the soft partition is within $1\%$ of
the oracle only above about $30$\,dB peak SNR; its nMSE is $2.3$ times the
oracle at $10$\,dB and about $60$ times at $-10$\,dB.  Multiplying by
\eqref{eq:shrinkage} restores \eqref{eq:mmse_est}, but only given the mode amplitudes,
bandwidths and noise floor.
\end{remark}

\subsection{Reconstruction-preserving allocation versus denoising
estimator}
\label{subsec:recon_vs_denoise}

The soft mask $s_k^*$ and the finite-noise MMSE estimator
$\hat U_k^\mathrm{MMSE}$ solve different
problems, and both differ from VMD:
\begin{itemize}[leftmargin=*]
\item $s_k^*$ is a \emph{reconstruction-preserving relative
  allocation rule}: by $\sum_k s_k^*=1$ (Theorem~\ref{thm:soft_opt}),
  \begin{equation}
    \sum_k s_k^*(\omega)\,X(\omega)\;=\;X(\omega),
  \label{eq:partition_recon}
  \end{equation}
  redistributing the signal without attenuating it.
\item $\hat U_k^\mathrm{MMSE}$ is a \emph{denoising estimator}:
  \begin{equation}
    \sum_k\hat U_k^\mathrm{MMSE}
    \;=\;\gamma\,X,
  \label{eq:partition_mmse}
  \end{equation}
  with $\gamma$ the shrinkage factor~\eqref{eq:shrinkage}, $0\le\gamma<1$ for
  $\sigma_N^2>0$: a real multiple of $X$, attenuating
  regions where noise dominates.
\item VMD~\eqref{eq:vmd_obj}, solved by ADMM, is a
  third problem: its fixed points (Section~\ref{sec:background}) are the
  inverse-square allocation~\eqref{eq:vmd_wiener} when a generally non-zero
  multiplier enforces reconstruction, and a shrunk, non-reconstructing inverse-square
  form otherwise; neither is~\eqref{eq:soft_opt} at finite $\alpha$ or
  the MMSE estimator~\eqref{eq:mmse_est}.
\end{itemize}
The two coincide as $\sigma_N^2\to0$ (Theorem~\ref{thm:mmse}). At finite noise,
the mask suits exact reconstruction of well-resolved modes, the MMSE estimator
inter-mode denoising; Remark~\ref{rem:finite_noise} gives
the analytic factor converting between them.

\subsection{Physical Interpretation of \texorpdfstring{$\alpha$}{alpha}}

By Theorem~\ref{thm:mmse}, $\alpha$ in~\eqref{eq:soft_opt}
encodes a prior on mode bandwidth.  This concerns the mask only: the VMD ADMM penalty cancels from the allocation at
a fixed point (\eqref{eq:admm_fp_identity}--\eqref{eq:vmd_wiener}), so with the
multiplier enforcing reconstruction and the centres given, VMD reaches the same
allocation for every $\alpha$.  Specifically,
\begin{equation}
  \alpha = \frac{1}{2\beta^2} \quad\Leftrightarrow\quad
  \beta = \frac{1}{\sqrt{2\alpha}},
\end{equation}
with $\beta$ the Lorentzian half-bandwidth (Table~\ref{tab:alpha_prior}).

\begin{table}[t]
\centering
\caption{Bayesian interpretation of the bandwidth parameter $\alpha$.
At fixed peak $\sigma_k^2$, \eqref{eq:lorentz_psd} (area $\pi\sigma_k^2\beta_k$ over $\mathbb{R}$)
tends pointwise to zero away from the centre as $\beta_k\to0$: a
vanishing-width prior, a delta only after renormalising the
area.}
\label{tab:alpha_prior}
\begin{tabular}{@{}ll >{\raggedright\arraybackslash}p{6.5cm}@{}}
\toprule
$\alpha$ & Prior $\beta$ & Mode extraction \\
\midrule
$\to\infty$ & $\to 0$ (vanishing width) & Inverse-square soft, $s_k^*\!\to\! d_k^{-2}/\!\sum_j d_j^{-2}$; VMD ADMM fixed point if $\tau>0$ enforces reconstruction (Rem.~\ref{rem:admm_fp}) \\
finite $\alpha$ & $1/\sqrt{2\alpha}$ & Lorentzian mask \eqref{eq:wiener_weights} \\
small & wide-band & Diffuse soft partition \\
\bottomrule
\end{tabular}
\end{table}

The hard Voronoi partition is not a limit of this family
(Remark~\ref{rem:not_voronoi}).
Choosing $\alpha$ thus specifies a prior on mode bandwidth rather than
tuning an algorithm; when unknown, it can be estimated from each band's FWHM:

\begin{corollary}[FWHM Estimator]
\label{cor:fwhm}
A Lorentzian of half-bandwidth $\beta$ has full width at
half maximum $\mathrm{FWHM} = 2\beta$, so $\alpha$ is estimated in closed form by
\begin{align}
  \hat{\alpha}_k &= \frac{2}{\mathrm{FWHM}_k^2}, \notag\\
  \mathrm{FWHM}_k &= \#\bigl\{\omega\in[b_{0,k},b_{1,k}]:
  P(\omega)\geq P_k^\mathrm{peak}/2\bigr\}\cdot\Delta\omega.
\label{eq:fwhm_est}
\end{align}
Here $[b_{0,k},b_{1,k}]$ is the Voronoi band of mode $k$ from
DAMD, $P_k^\mathrm{peak}=\max_{\omega\in[b_{0,k},b_{1,k}]} P(\omega)$
its peak power, and $\Delta\omega$ the
uniform frequency resolution.  Accuracy depends on
$n=2\beta/\Delta\omega$ (bins across the FWHM), not on $\alpha$:
the bin count is within one bin of $n$, so
$|\hat\alpha_k/\alpha_k-1|\le(1-1/n)^{-2}-1\approx2/n$
in the worst case over the centre's grid position and, averaged over it, to first
order in $1/n$, $4f(1-f)/n\le1/n$ ($f$ the
fractional part of $n$); without noise, a relative error below $10\%$ is guaranteed for
$n\geq22$.  Under noise the spectrum must be smoothed or averaged, which is
necessary but not sufficient.  The band must also isolate the mode: the estimate degrades once a neighbour of
comparable amplitude lies within about three FWHM, whereas a centre error is harmless
while the band contains the half-power region (Appendix~C).

The Type-II maximum likelihood estimate
$\alpha^*=\arg\max_\alpha \log p(X|\{\omega_k\},\alpha)$ has gradient
\begin{equation}
\begin{aligned}
  \frac{\partial\log p}{\partial\alpha}
  &= \sum_\omega \Bigl[
      \frac{P(\omega)}{S_X^2(\omega)} - \frac{1}{S_X(\omega)}
    \Bigr]
    \frac{\partial S_X}{\partial\alpha},\\
  \frac{\partial S_X}{\partial\alpha}
  &= -2\sigma^2 \sum_k
    \frac{(\omega-\omega_k)^2}{[1+2\alpha(\omega-\omega_k)^2]^2}.
\end{aligned}
\label{eq:mle_grad}
\end{equation}
The FWHM estimator~\eqref{eq:fwhm_est} is closed-form and can initialise \eqref{eq:mle_grad}.
For non-Lorentzian modes $\hat\alpha_k$ is a \emph{shape-matched} bandwidth parameter from the empirical half-width, not an estimate of a true Lorentzian $\alpha$, and the Bayesian reading is approximate.
\end{corollary}

\section{Centre Update and Convergence}
\label{sec:convergence}

\subsection{Soft Centre Update (M-step)}

Given $s_k^*$ (Theorem~\ref{thm:soft_opt}), the
centres minimise $\tilde J_\alpha$ (Definition~\ref{def:soft_vmd}):

\begin{proposition}[Soft M-step]
\label{prop:mstep}
For fixed soft weights $\{s_k^*(\omega;\alpha)\}$, the objective
$\tilde J_\alpha$ is minimised by the weighted centroids
\begin{equation}
  \omega_k^* = \frac{\int [s_k^*(\omega)]^2
    P(\omega)\,\omega\,d\omega}
    {\int [s_k^*(\omega)]^2 P(\omega)\,d\omega}.
\label{eq:mstep}
\end{equation}
These also minimise $J_\mathrm{VMD}$, since the regulariser
$\tfrac{1}{2\alpha}\int\sum_k s_k^2 P\,d\omega$ does not depend on
$\{\omega_k\}$; \eqref{eq:mstep} is
the VMD frequency update~\eqref{eq:admm_freq} with $\hat U_k$
replaced by the soft extraction.
\end{proposition}

\emph{Proof:} see Appendix~\ref{app:proofs}.

\subsection{Convergence}

\begin{proposition}[Monotone decrease of the objective]
\label{prop:conv}
The iteration
\begin{align}
  \text{E-step:}\quad &s_k^{(n+1)} = \arg\min_{s\in\Delta^K}
    \tilde J_\alpha(s;\{\omega_k^{(n)}\}),\label{eq:estep}\\
  \text{M-step:}\quad &\omega_k^{(n+1)} = \arg\min_{\omega_k}
    \tilde J_\alpha(s^{(n+1)};\{\omega_k\}),\label{eq:mstep_iter}
\end{align}
produces a
non-increasing sequence $\tilde J_\alpha^{(n)}$; since $\tilde J_\alpha\geq 0$, the
objective values converge by the bounded monotone sequence
theorem.
\end{proposition}

\emph{Proof:} see Appendix~\ref{app:proofs}.

\begin{remark}[Scope of Proposition~\ref{prop:conv}]
It concerns $\tilde J_\alpha^{(n)}$ only, not stationarity, global optimality
or convergence of the iterates (neither established nor excluded).  If two centres coincide then $s_i\equiv s_j$ by
\eqref{eq:soft_opt} and the M-step returns $\omega_i'=\omega_j'$, so coincidence is an
invariant set: limit points depend on the initialisation and stationary configurations
are not unique.
\end{remark}

\begin{remark}[E-step uses $\tilde J_\alpha$, not $J_\mathrm{VMD}$]
\label{rem:estep_obj}
The E-step minimises $\tilde J_\alpha$, not
$J_\mathrm{VMD}=\sum_k\int|\omega-\omega_k|^2 s_k^2 P\,d\omega$, whose minimiser over
$\Delta^K$ where $P(\omega)>0$ is the
inverse-square allocation
\begin{equation}
  s_k(\omega)=\frac{|\omega-\omega_k|^{-2}}{\sum_j|\omega-\omega_j|^{-2}},
\label{eq:unreg_min}
\end{equation}
This is
\eqref{eq:vmd_wiener} and \eqref{eq:invsq_limit}: the unregularised
functional, the $\alpha\to\infty$ Lorentzian endpoint and the VMD ADMM
fixed point coincide.  The $1/(2\alpha)$ regulariser
moves the minimiser into the simplex
interior at the Lorentzian point (Theorem~\ref{thm:soft_opt});
without it the problem is degenerate at
$\omega=\omega_k$.
\end{remark}

\begin{remark}[Relation to VMD ADMM fixed points]
\label{rem:admm_fp}
By \eqref{eq:admm_fp_identity}, every fixed point of the VMD ADMM mode update
satisfies $\hat U_k\propto d_k^{-2}$, and is \eqref{eq:vmd_wiener} when the dual
variable enforces reconstruction; by
Corollary~\ref{cor:hard_limit} this is
$\lim_{\alpha\to\infty}s_k^*(\omega;\alpha)$.  The Lorentzian family thus
\emph{meets} the VMD fixed point at its $\alpha\to\infty$ endpoint, and besides the
prior bandwidth $\beta=1/\sqrt{2\alpha}$ (Section~\ref{sec:mmse}), $\alpha$ controls how closely
the allocation approaches it.
Fixed points of \eqref{eq:estep}--\eqref{eq:mstep_iter} at finite $\alpha$ are
\emph{not} ADMM fixed points; they agree only in the limit.
\end{remark}

\begin{remark}[Practical purpose of the convergence analysis]
Proposition~\ref{prop:conv} shows that the alternating iteration decreases the
objective monotonically, and Remark~\ref{prop:rate} estimates its sensitivity; with accurate detected
centres~\cite{jia2025damd}, evaluating~\eqref{eq:soft_opt} at them
suffices, and Algorithm~\ref{alg:damd_soft} needs no EM iteration.
\end{remark}

\begin{remark}[Heuristic sensitivity estimate]
\label{prop:rate}
A sensitivity calculation, not a theorem: no contraction or rate is claimed,
and the spectral-radius argument (with the signs and cancellations of the
Jacobian entries) is omitted, since Algorithm~\ref{alg:damd_soft} does not iterate.
Let $\Delta\omega_{\min} = \min_{j\neq k}|\omega_j^*-\omega_k^*|$ be
the minimum separation at a fixed point of
\eqref{eq:estep}--\eqref{eq:mstep_iter}, and assume
$\alpha\Delta\omega_{\min}^2 \gg 1$, with
(i)~distinct centres; (ii)~every M-step denominator $\int[s_k^*]^2P\,d\omega$
bounded away from zero; and (iii)~the part of $P$ far from all centres, where the
weights do not localise, small relative to modal energies. For fixed $K$, $\alpha$ and $P$, the Jacobian entries of the composed map
(Appendix~\ref{app:proofs}) may be small when the spectral power is
concentrated near well-separated centres; this does not establish monotonicity,
contraction or a convergence rate.
\end{remark}

\section{Algorithm and Complexity}
\label{sec:algorithm}

\subsection{Algorithm}

Algorithm~\ref{alg:damd_soft} summarises the method.  Its two steps are
decoupled: Step~1 (detection) produces centre frequencies by any
method (density-based clustering, VMD ADMM, manual specification or
domain-specific algorithms), and Step~2 (extraction,
Theorems~\ref{thm:soft_opt}--\ref{thm:mmse}) computes the modes in closed
form from $\{\omega_k\}$ and $\alpha$ alone, so switching from hard to soft
extraction replaces $\mathbf{1}_{B_k}(\omega)$ by $s_k^*(\omega;\alpha)$
without recomputing detection.

\begin{algorithm}[t]
\caption{DAMD with Soft Spectral Partition}
\label{alg:damd_soft}
\begin{algorithmic}[1]
\REQUIRE Time-frequency map $\mathcal{T}(\omega,t)$,
         penalty $\alpha$ (or estimated via Corollary~\ref{cor:fwhm}),
         extraction mode $\in\{\texttt{hard},\texttt{soft}\}$
\ENSURE  Complex mode functions $\{U_k(\omega,t)\}$ per frame;
         waveforms follow by inverse transform for invertible
         $\mathcal{T}$ only after associating indices $k$
         across frames, which is not done here
         (Section~\ref{sec:conclusion})

\FOR{$t = 1$ to $T$}
  \STATE \textbf{[Step 1 -- Mode detection (DAMD, algorithm-agnostic)]}
  \STATE Extract power spectrum $P(\omega) = |\mathcal{T}(\omega,t)|^2$
  \STATE Density-based clustering on $P(\omega)$ $\to$
         centres $\{\omega_{k,t}^*\}$, $K_t^*$
         \hfill $O(S\cdot N_\mathrm{M}\cdot F)$

  \STATE \textbf{[Step 2 -- Mode extraction (this paper)]}
  \IF{extraction mode $=$ \texttt{hard}}
    \STATE $U_k(\omega,t) \gets \mathbf{1}_{B_k}(\omega)\cdot\mathcal{T}(\omega,t)$
           \hfill $O(F\cdot K_t)$
  \ELSE[\texttt{soft}]
    \IF{$\alpha$ not specified}
      \STATE $\hat\alpha_k \gets 2/\mathrm{FWHM}_{k,t}^2$
             (Corollary~\ref{cor:fwhm}, on a smoothed spectrum) \hfill $O(F\cdot K_t)$
    \ENDIF
    \FOR{$k = 1$ to $K_t^*$}
      \STATE $s_k^*(\omega) \gets
             \dfrac{[1{+}2\alpha_k|\omega{-}\omega_k|^2]^{-1}}
                        {\sum\nolimits_j[1{+}2\alpha_j|\omega{-}\omega_j|^2]^{-1}}$
             \hfill $O(F)$
      \STATE $U_k(\omega,t) \gets s_k^*(\omega)\cdot\mathcal{T}(\omega,t)$
             \hfill $O(F)$
    \ENDFOR
  \ENDIF
\ENDFOR
\end{algorithmic}
\end{algorithm}

Step~1 is published DAMD \cite{jia2025damd}; Step~2, our practical
contribution, adds non-iterative Lorentzian extraction at $O(F\cdot K_t)$ per frame.
The inverse-square allocation~\eqref{eq:vmd_wiener} can be evaluated just as
directly from the same centres and already gives most of the gain over hard extraction
(Experiments~C and~C$'$); finite $\alpha$ adds a smaller gain for close modes,
together with a well-posed objective (Theorem~\ref{thm:soft_opt}; the unregularised
one, minimised by~\eqref{eq:vmd_wiener}, is degenerate at the centres,
Remark~\ref{rem:estep_obj}) and an MMSE reading in which $\alpha$ is a prior
bandwidth (Theorem~\ref{thm:mmse}).
The soft weight of Step~2 is \eqref{eq:weighted_opt} at equal amplitudes; with the
per-mode $\hat\alpha_k$ of Corollary~\ref{cor:fwhm} it differs from
$[|\omega-\omega_k|^2+1/(2\alpha_k)]^{-1}$ by a factor $\alpha_k$, and only the form
above is the minimiser (Remark~\ref{rem:weighted_fix}).  The soft E-M
iteration of Section~\ref{sec:convergence} (related to the VMD ADMM fixed
point only as $\alpha\to\infty$, Remark~\ref{rem:admm_fp}) is a theoretical
consistency check, not a practical recommendation.

\begin{remark}[Time-domain reconstruction]
\label{rem:timedomain}
For $\hat{U}_k(\omega,t) =
s_k^*(\omega)\cdot\mathcal{T}(\omega,t)$, first, $s_k^*(\omega)$ is a global minimiser of
the regularised soft VMD problem at fixed centres, unique on the support of $P$
(Theorem~\ref{thm:soft_opt}).
Second, under the Lorentzian mode model
(Definition~\ref{def:lorentz_model}) with equal $\sigma_k^2$ and $\alpha_k$, and in
the low-noise limit of Theorem~\ref{thm:mmse}, $\hat{U}_k(\omega,t)$ coincides
with the MMSE estimator of the \emph{complex} mode coefficient
$U_k^\mathrm{true}(\omega,t)$ (at finite noise the two differ by the shrinkage
factor~\eqref{eq:shrinkage}): the
cross-spectrum $S_{U_k^\mathrm{true},X}(\omega) = S_k(\omega)$ is
real, so the Wiener--Hopf formula $\hat{U}_k = (S_k/S_X)\,X$ applies the
real weight $s_k^*(\omega)$ to $\mathcal{T}(\omega,t)$ and preserves its phase.
Waveforms $\hat{u}_k(n)$ follow by inverting the transform (STFT pair specified in
Appendix~\ref{app:stft}).
Reconstruction is the weaker statement: since
$\sum_k s_k^*=1$ pointwise, linearity of synthesis gives
$\sum_k\mathcal{T}^{-1}[s_k^*\mathcal{T}x]=\mathcal{T}^{-1}[\mathcal{T}x]$ for every
invertible transform.  For the unit-hop STFT the round trip is exact to machine
precision (relative error $\sim6\times10^{-16}$); the nonlinear synchrosqueezed transforms of
Experiment~D are never inverted, so no result is affected.  Transferring per-bin MMSE optimality to the waveform
would also require the masked array to stay in the range of the analysis operator, which
it does not (residuals in Appendix~G).  Synthesis is a bounded linear operator, so the
waveform error is bounded by a constant times the coefficient-domain error, but
Theorem~\ref{thm:mmse} is a coefficient-domain result and we make no waveform-domain
MMSE claim.

\end{remark}

\subsection{Complexity Comparison}

The principal baseline for time-varying signals, STVMD, applies
VMD independently at each frame:
\begin{equation}
  \mathcal{C}_\mathrm{STVMD}
  = O\!\left(T\cdot F\log F\right)
  + O\!\left(T\cdot K\cdot N_\mathrm{A}\cdot F\right).
\end{equation}
Table~\ref{tab:complexity} compares the three methods.

\begin{table}[t]
\centering
\caption{Per-frame complexity and exactness.
$\checkmark$: exact minimiser of the regularised objective~\eqref{eq:soft_vmd} (closed form, one pass);
$\times$: iterative ADMM, whose fixed points ($\tau>0$) are~\eqref{eq:vmd_wiener}, not~\eqref{eq:soft_opt}.
STVMD: short-time VMD; DAMD: hard extraction; DAMD-Soft: soft extraction
(Algorithm~\ref{alg:damd_soft}).}
\label{tab:complexity}
\begin{tabular}{lccc}
\toprule
 & STVMD & DAMD & DAMD-Soft \\
\midrule
$K$             & fixed  & auto         & auto \\
Transform ($1\times$) & $O(F\!\log\!F)$ & $O(F\!\log\!F)$ & $O(F\!\log\!F)$ \\
Mode detection  & ---    & $O(S N_\mathrm{M} F)$ & $O(S N_\mathrm{M} F)$ \\
Extraction      & $O(K N_\mathrm{A} F)$ & $O(FK)$ & $O(FK)$ \\
Soft weights    & inside ADMM & no      & $O(FK)$ \\
Exact weights   & $\times$ & ---        & $\checkmark$ \\
\bottomrule
\end{tabular}
\begin{tablenotes}
\small
\item $N_\mathrm{A}$: ADMM iters (200--500); $N_\mathrm{M}$: mean-shift iters per seed ($13.6$--$16.5$ on average, see text); $S\leq F$: seeds.
  The transform ($\log F$) is a one-time per-frame cost; the ADMM loop has
  elementwise frequency-domain updates and no per-iteration FFT.
\end{tablenotes}
\end{table}

Soft extraction adds a fraction
$K/(S\cdot N_\mathrm{M})$ of the mean-shift cost in operation counts.  In the setting of Experiment~S5,
$S=13$ seeds per frame take $N_\mathrm{M}=13.6$--$16.5$ iterations on average (the
slowest seed of a frame $30$--$35$ in the median, up to about $300$), so with the
detected $K_t$ the fraction is $3$--$5\%$.
  Measured wall-clock time is larger,
$9$--$11\%$ of the DAMD pipeline, owing to different constant factors.

The key gain is in extraction: DAMD-Soft replaces STVMD's iterative ADMM
($O(K\cdot N_\mathrm{A}\cdot F)$ per frame) by one closed-form pass
($O(F\cdot K)$), removing the $O(N_\mathrm{A})$ factor; detection needs far fewer iterations
(Table~\ref{tab:complexity}). Measured end to end, the ratio rises from $236\times$ at $N=64$ to $390\times$ at
$N=2048$ on a three-component non-stationary signal (Experiment~S5; timings, dispersion and setup in
Appendix~\ref{app:runtime}).  The ratio depends on $N$, and the methods use different mode-selection rules: STVMD is given
$K=10$ whereas DAMD returns $K_t\in[5,11]$, mean $8.0$.  The ratio is thus a pipeline comparison, including mode-number
determination, rather than a speed-up for an identical decomposition.  The outputs also differ: STVMD
returns waveforms and its timing includes the inverse transform, whereas the
DAMD pipeline, soft stage included, is timed on coefficients.  Settings are each method's defaults, and accuracy
is not compared.  Without label association across
frames the outputs cannot be scored mode by mode, but both must sum back to the signal,
which the soft allocation does to $3.7\times10^{-16}$ (the
transform's round-trip error) and STVMD, with its default dual step, to
$9.1\times10^{-3}$.

\section{Experiments}
\label{sec:experiments}

Table~\ref{tab:exp_map} maps experiments to statements.  Experiments follow the argument chain:
the VMD fixed point (Remark~\ref{rem:admm_fp}), the partition's properties in
theorem order, then real signals; the timing comparison of
Section~\ref{sec:algorithm} is Experiment~S5 (Appendix~\ref{app:runtime}).  Experiment~D,
the consistency check~S1 and the pipeline experiments~S4 and~S5 use DAMD~\cite{jia2025damd}
with $\alpha=2000$; Experiments~A--C$'$ do not run DAMD (A instruments the VMD ADMM, B is
analytic, C and~C$'$ are Monte Carlo with $\alpha=5000$), each stating its own setting.

\textbf{Reproducibility.}  Random draws use fixed seeds ($42$ for
Experiments~A, B, C, C$'$, S4 and S5; $0$ for S1--S3), so the numbers below are
exactly reproducible (S5, a wall-clock timing, only up to the machine).  The
four third-party recordings of Experiment~D are identified by record and channel in
Section~\ref{subsec:exp_f}.

\begin{table}[t]
\centering
\caption{Experiment--theory mapping.  Rows prefixed S are in
Appendices~\ref{app:exp_b}--\ref{app:runtime}.  Experiment~S1, an implementation unit test, is not evidence for any
statement and is not listed.}
\label{tab:exp_map}
\footnotesize
\setlength{\tabcolsep}{2pt}
\renewcommand{\arraystretch}{1.05}
\begin{tabular}{@{} l l l @{}}
\toprule
\textbf{Exp.} & \textbf{Statement tested} & \textbf{Nature of the test} \\
\midrule
A   & Rem.~\ref{rem:admm_fp}                                & Against the solver's output \\
B   & Cor.~\ref{cor:hard_limit}, Rem.~\ref{rem:not_voronoi} & Analytic, exact \\
C   & Thm.~\ref{thm:mmse}, Rem.~\ref{rem:finite_noise}      & Monte Carlo vs.\ oracle \\
C$'$ & Scope of Thm.~\ref{thm:mmse}                     & Monte Carlo under misspecification \\
D   & Boundary, allocation difference                & Descriptive, no ground truth \\
\midrule
 S2 & Cor.~\ref{cor:fwhm}       & Analytic and noisy spectra \\
S3 & Thm.~\ref{thm:soft_opt}   & Against a numerical minimiser \\
S4 & Scope of Thm.~\ref{thm:mmse}, time domain & Known components, full pipeline \\
S5 & Table~\ref{tab:complexity}  & Timing, not task-matched \\
\bottomrule
\end{tabular}
\end{table}

\subsection{Experiment A: Where the Lorentzian Family Meets the VMD Fixed Point}
\label{subsec:exp_a}

\textit{Purpose}: Test Remark~\ref{rem:admm_fp} on the ADMM solver's own mode
spectra, not on a second evaluation of a closed form.

\textit{Setup}: $K=3$ AM-modulated tones at $110$, $200$ and $330$\,Hz
(envelopes between $0.4$ and $1.0$) in white Gaussian noise of standard
deviation $0.05$ ($f_s=1000$\,Hz, $N=2048$, seed $42$, mirror extension as in
\cite{dragomiretskiy2014vmd}, $\alpha=2000$ in the convention of
\eqref{eq:soft_vmd}).  An instrumented ADMM with an active dual variable
(step $\tau=2$) and centres initialised at the true frequencies retains its mode
spectra $\hat U_k(\omega)$.  Since \eqref{eq:admm_fp_identity} characterises \emph{fixed points},
which require a vanishing primal residual (otherwise the multiplier keeps changing),
the iteration stops only when $\|\sum_k\hat U_k-X\|_2/\|X\|_2<10^{-10}$, not on
the relative change of $\hat U_k$, which is met long before.
The realised mask $|\hat U_k(\omega)|/|X(\omega)|$ is compared with the inverse-square
limit \eqref{eq:vmd_wiener} and the Lorentzian \eqref{eq:wiener_weights} at the
converged centres, and
$\|\hat U_k-\hat U_k^{\mathrm{closed}}\|_2/\|\hat U_k\|_2$ is recorded for $\alpha$
from $10^2$ to $3.2\times10^5$; beyond that the tolerance is not reached within
$4\times10^5$ iterations, the dual updates needed growing in proportion to
$\alpha$.

\begin{figure*}[t]
\centering
\includegraphics[width=\textwidth]{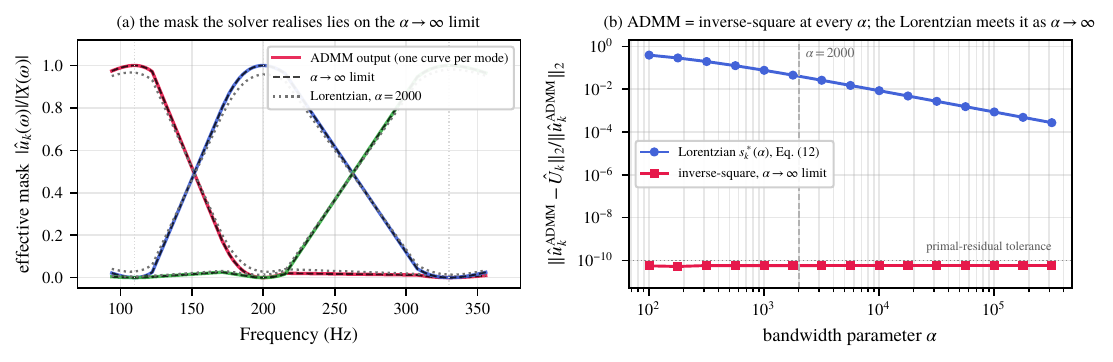}
\caption{Experiment~A (Remark~\ref{rem:admm_fp}).  (a) Mask realised by the ADMM
solver, from its own output.  (b) Distance from the converged ADMM spectra to each
closed form versus $\alpha$.}
\label{fig:exp_a}
\end{figure*}

\textit{Result}: The solver's mask lies on the inverse-square limit
(Fig.~\ref{fig:exp_a}(a)).  At $\alpha=2000$ the tolerance is reached after $1367$
iterations with primal residual $9.9\times10^{-11}$ and
$\|\hat\lambda\|_2/\|X\|_2 = 3.70$; the converged centres are
$\omega_k = (109.74,\; 199.84,\; 330.47)$\,Hz, and the per-mode distances
$\|\hat U_k - s_k^{\mathrm{inv}}X\|_2/\|\hat U_k\|_2$ lie between $2\times10^{-11}$
and $1\times10^{-10}$, i.e.\ at the convergence tolerance, against
$(0.045,\;0.053,\;0.024)$ to the Lorentzian.  Across the sweep the inverse-square
distance stays at tolerance at \emph{every} $\alpha$, whereas the Lorentzian
distance decreases as $O(1/\alpha)$ (Fig.~\ref{fig:exp_a}(b)): by
\eqref{eq:admm_fp_identity}, $\alpha$ cancels from the fixed-point allocation once
$R=0$, so it is the Lorentzian family that approaches the fixed point.
The relative change of $\hat U_k$ over the final iteration, the iterate
change on which both the stopping rule of~\cite{dragomiretskiy2014vmd} and the ADMM
dual residual of~\cite{boyd2011distributed} are built, is
$9.5\times10^{-13}$.  The multiplier does not vanish but equals the fixed-point value $\hat\lambda=4\alpha\hat X/\sum_jd_j^{-2}$ from \eqref{eq:admm_fp_identity}
with $R=0$, to a relative error of $4\times10^{-9}$; the distance
$\|\hat U_k-s_k^{\mathrm{inv}}X\|$ reaches tolerance precisely when $\hat\lambda$ does.
With the dual step off ($\tau=0$, so $\hat\lambda\equiv0$, $2000$ iterations) the
solver instead converges to the shrunk inverse-square form of
Section~\ref{sec:background}, to within $7\times10^{-16}$, leaving a reconstruction
residual of $9.7\times10^{-2}$.

\emph{Experiment~S1}, a consistency check between \eqref{eq:soft_opt} and an
independent evaluation of \eqref{eq:wiener_weights}, is a unit test rather than
evidence (Appendix~\ref{app:exp_b}).

\subsection{Experiment B: Limits of the Lorentzian Family}
\label{subsec:exp_c}

\textit{Purpose}: Verify Corollary~\ref{cor:hard_limit} and
Remark~\ref{rem:not_voronoi}: as $\alpha\to\infty$ the Lorentzian partition tends to
the inverse-square allocation $d_k^{-2}/\sum_j d_j^{-2}$, \emph{not} the hard Voronoi
indicator, which lies outside the family.

\begin{figure*}[t]
\centering
\includegraphics[width=\textwidth]{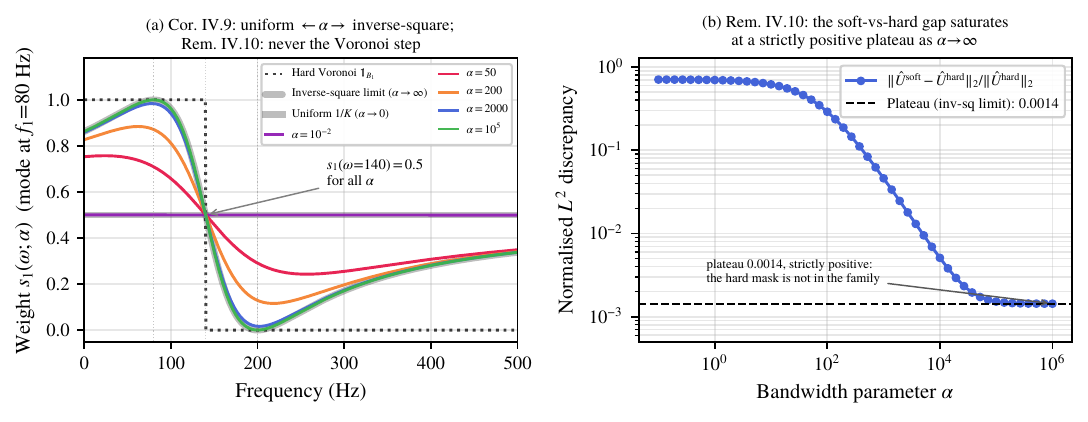}
\caption{Experiment~B, mode at 80\,Hz.  (a) Lorentzian weights $s_k(\omega;\alpha)$
for $\alpha\in\{10^{-2},\,50,200,2000,10^5\}$ ($10^{-2}$ is the uniform
$\alpha\to0$ endpoint), with the inverse-square asymptote
$d_k^{-2}/\!\sum_j d_j^{-2}$ (dashed) and the Voronoi indicator $\mathbf{1}_{B_k}$
(dotted).  (b) Soft--hard extraction discrepancy versus $\alpha$
(Cor.~\ref{cor:hard_limit}, Rem.~\ref{rem:not_voronoi}).}
\label{fig:exp_c}
\end{figure*}

\textit{Setup}: Two tones ($f_1=80$\,Hz, $f_2=200$\,Hz, $f_s=1000$\,Hz, $K=2$).
Fig.~\ref{fig:exp_c}(a) compares $s_k(\omega)$ with the asymptote
$d_k^{-2}/(d_k^{-2}+d_j^{-2})$ and $\mathbf{1}_{B_k}$; Fig.~\ref{fig:exp_c}(b) shows
$\|\hat U_k^{\mathrm{soft}}(\alpha)-\hat U_k^{\mathrm{hard}}\|_2/
 \|\hat U_k^{\mathrm{hard}}\|_2$ over six decades of $\alpha$.

\textit{Result}: The Lorentzian profile converges to the inverse-square S-curve,
not the Voronoi step; at the midpoint $\omega=140$\,Hz its weight is $0.5$ for every $\alpha$, where the indicator
jumps from~$1$ to~$0$.  The soft--hard discrepancy saturates at a plateau set by
the inverse-square mismatch, small but strictly positive for narrow-band tones.

\subsection{Experiment C: MMSE Optimality}
\label{subsec:exp_d}

\textit{Purpose}: Test Theorem~\ref{thm:mmse} in its low-noise limit and
Remark~\ref{rem:finite_noise} outside it, and compare soft with hard extraction.

\textit{Setup}: Frequency-domain Monte Carlo with $K=2$ Lorentzian modes
(normalised centres $\omega_1=0.12$, $\omega_2=0.30$, $\alpha=5000$,
$\sigma_1^2=\sigma_2^2=1$, $F=512$ bins, $N=2000$ trials per SNR).  Per bin and trial,
$U_k^{\mathrm{true}}(\omega)\sim\mathcal{CN}(0,S_k(\omega))$ and
$N(\omega)\sim\mathcal{CN}(0,\sigma_N^2)$ are drawn independently, and
$X=\sum_kU_k^{\mathrm{true}}+N$.  Extractors: hard (Voronoi), soft
($\alpha=\alpha_\mathrm{true}$) and oracle MMSE ($S_k/S_X$).
The SNR is $\sigma^2/\sigma_N^2$, peak mode spectral density over noise density
per bin.
The metric for mode~1 is
$\mathrm{nMSE} = \mathbb{E}\bigl[\sum_\omega|\hat U_1(\omega)-U_1^{\mathrm{true}}(\omega)|^2\bigr]\big/
                 \mathbb{E}\bigl[\sum_\omega|U_1^{\mathrm{true}}(\omega)|^2\bigr]$,
estimated by the sample mean over trials.

\begin{figure*}[t]
\centering
\includegraphics[width=\textwidth]{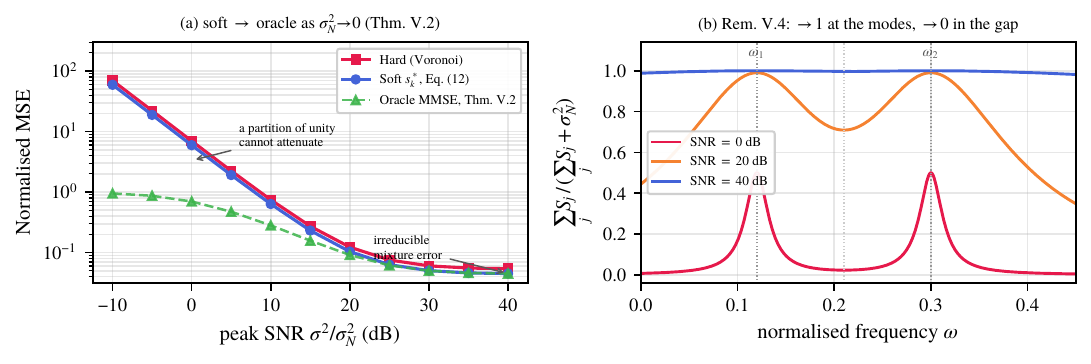}
\caption{Experiment~C (Theorem~\ref{thm:mmse}, Remark~\ref{rem:finite_noise}).
(a) nMSE versus peak SNR; relative Monte Carlo standard errors are below $0.3\%$,
smaller than the markers.  (b) Shrinkage factor \eqref{eq:shrinkage} across frequency
at three SNRs.}
\label{fig:exp_d}
\end{figure*}

\begin{figure*}[t]
\centering
\includegraphics[width=\textwidth]{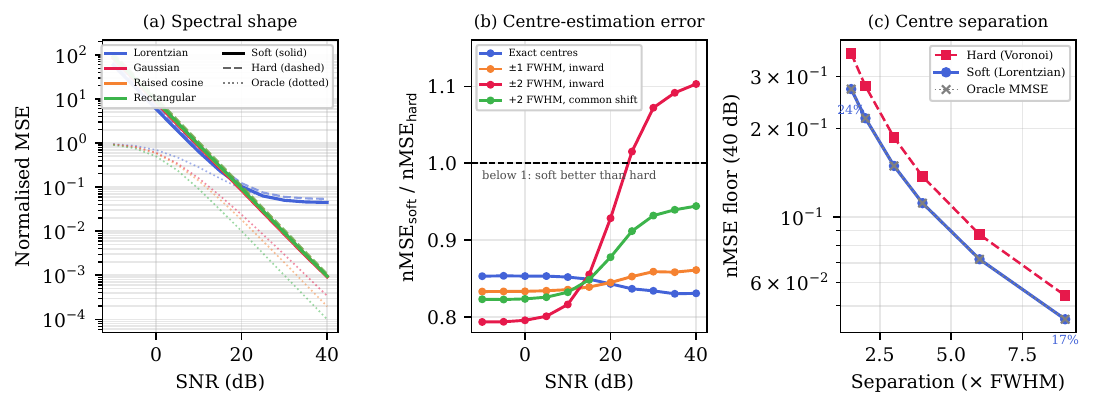}
\caption{Experiment~C$'$.  (a)~nMSE versus SNR for four FWHM-matched spectral
shapes; solid: soft (Lorentzian), dashed: hard (Voronoi), dotted: oracle MMSE.
(b)~Centre error (Lorentzian modes): $\mathrm{nMSE}_\mathrm{soft}/\mathrm{nMSE}_\mathrm{hard}$
with the true modes fixed and the assumed centres displaced; values below one favour
the soft mask, and a symmetric displacement of $2$ FWHM reverses the ranking from
$25$\,dB on.  (c)~nMSE at $40$\,dB versus centre separation (Lorentzian modes, exact
centres); circles: soft, squares: hard.}
\label{fig:robustness}
\end{figure*}

\textit{Result}: The soft partition approaches the oracle as $\sigma_N^2\to0$, as
Theorem~\ref{thm:mmse} states: they differ by less than $0.1\%$ at $40$\,dB and by
$0.4\%$ at $30$\,dB.  Their common value, $\mathrm{nMSE}\approx0.045$, is the
irreducible error of separating two overlapping Lorentzian modes, a property of the
model rather than a numerical floor.  At lower SNR they diverge (the soft nMSE is
$2.3$ times the oracle at $10$\,dB and about $60$ times at $-10$\,dB), because
$\sum_ks_k^*=1$ prevents any attenuation.  The shrinkage factor \eqref{eq:shrinkage}
closes this gap by construction ($s_k^*$ times it is algebraically $S_k/S_X$), so it
is not plotted.  Hard extraction stays above the soft
curve at every SNR tested: in the low-noise limit by Theorem~\ref{thm:mmse}, the soft
mask then being the MMSE estimator; at finite noise empirically, not by
Corollary~\ref{cor:strict}, which concerns the objectives \eqref{eq:Jvmd} and \eqref{eq:Jkm} rather than
mean-square error.
The VMD fixed-point allocation~\eqref{eq:vmd_wiener} (Experiment~A), evaluated on
the same draws and centres, is within $0.5\%$ of the soft mask in nMSE at every SNR of
Fig.~\ref{fig:exp_d}, not plotted ($0.0450$ for both at $40$\,dB, against $0.0541$ for hard
extraction).  For closer modes (Fig.~\ref{fig:robustness}(c)), its nMSE exceeds the
soft mask's by $0.8\%$, $2.0\%$ and $3.7\%$ at $3$, $2$ and $1.5$ FWHM at $40$\,dB
($0.1503$, $0.2206$ and $0.2819$ against $0.1492$, $0.2164$ and $0.2719$; percentages
from unrounded values), and by $1.5$--$3.3\%$ at $10$\,dB, whereas hard extraction
exceeds it by $25$--$32\%$ at $40$\,dB and by $14$--$16\%$ at $10$\,dB.  The VMD
allocation thus captures most of the advantage over hard extraction; finite $\alpha$
adds a few per cent for close modes.

\textbf{Experiment~C$'$: Robustness to spectral model misspecification.}
In Experiment~C the soft mask matches the data by construction
(Definition~\ref{def:lorentz_model}).  We repeat it ($2000$ trials per SNR) with Gaussian, raised-cosine and rectangular spectra, each FWHM-matched to
the Lorentzian baseline ($\beta=1/\sqrt{2\alpha}\approx0.01$); the soft mask is always
the Lorentzian~\eqref{eq:soft_opt} with $\alpha=5000$, and the oracle is $S_k/S_X$
with the true PSD.  Standard errors are below $0.3\%$.

Under all four shapes the soft mask beats hard extraction at every SNR tested ($-10$ to $40$\,dB) (Fig.~\ref{fig:robustness}(a)).  The
soft nMSE, within $0.1\%$ of the oracle's under the Lorentzian model, rises to roughly
three times the oracle's for the Gaussian and ten for the rectangular shape at 40\,dB: the MMSE
equivalence of Theorem~\ref{thm:mmse} is specific to the Lorentzian model, but the
advantage over hard extraction is preserved.

Fig.~\ref{fig:robustness}(b) isolates centre error at Lorentzian shape.  The true
modes stay at $\omega_1=0.12$ and $\omega_2=0.30$ and only the centres \emph{assumed}
by the masks are displaced, as by a detection error: towards each other by $1$ and by
$2$ FWHM, and both in the same direction by $2$ FWHM.  A symmetric displacement leaves
the Voronoi boundary, hence the hard mask, unchanged, so the ratio
$\mathrm{nMSE}_\mathrm{soft}/\mathrm{nMSE}_\mathrm{hard}$ is plotted.  At
$1$ FWHM the soft mask remains better at every SNR ($0.047$ against $0.054$ at
$40$\,dB).  At $2$ FWHM it is better only up to $20$\,dB and worse from $25$\,dB on
($0.060$ against $0.054$ at $40$\,dB): once the centre error is comparable to the mode
width, the smooth mask leaks more energy than the displaced hard boundary.  A common
displacement of $2$ FWHM, which moves the boundary too, leaves the soft mask better at
every SNR ($0.068$ against $0.072$).  With the non-Lorentzian shapes of panel~(a) and
an inward displacement of $1$ FWHM, the soft mask is better up to $30$\,dB but worse
from $35$\,dB on, with $40$\,dB nMSE about $1.6$--$1.9$ times that of hard extraction
($0.0019$ against $0.0010$ for the Gaussian shape).  The advantage is therefore conditional on centre accuracy: of the order of one FWHM when the modes
are Lorentzian, and finer than that at high SNR when they are not.  Whether a
particular detector achieves this is not established here.

\textbf{Mismatched $\alpha$.}  With the mask's $\alpha$ scaled by $\tfrac14$,
$4$, $16$ and $\tfrac1{16}$ relative to the true bandwidth, at exact centres, the
$40$\,dB nMSE is $0.0451$, $0.0450$, $0.0450$ and $0.0483$ against $0.0450$.  Only the
$16$-fold \emph{under}-estimate is visible, since scaling $\alpha$ barely changes the
weight ratios in \eqref{eq:soft_opt} where the modes are comparable.  At $2$ FWHM
separation the same four mismatches give $0.232$, $0.219$, $0.220$
and $0.315$ against $0.216$, the last worse than hard extraction ($0.278$).  A factor
of a few in $\alpha$ thus suffices, and the damaging direction is a prior that is too
wide.  Corollary~\ref{cor:fwhm} delivers such a factor without noise once the FWHM spans
a handful of bins, and under noise only on an averaged spectrum (Appendix~C).

Fig.~\ref{fig:robustness}(c) varies the separation from $1.5$ to $9$ FWHM (the
default of panels~(a) and~(b)), with exact centres and Lorentzian shape.  As spectral overlap
increases, the irreducible floor rises from $0.045$ at $9$ FWHM to $0.272$ at
$1.5$ FWHM.  The soft mask stays below the hard mask at every separation and SNR,
its advantage growing monotonically from $16.9\%$ to $24.3\%$ at $40$\,dB as the
modes approach: smooth allocation helps most when modes overlap.

\textbf{Frequency resolution.}  To include window leakage, the
modes are also generated as time series (complex first-order autoregressions with the
mask's half-width near their peaks) and analysed by a Hann-windowed STFT at $0.32$ to
$20.5$ bins across the FWHM.  The ratio
$\mathrm{nMSE}_\mathrm{soft}/\mathrm{nMSE}_\mathrm{hard}$ is $0.79$, $0.74$ and $0.71$
at $40$, $20$ and $0$\,dB whenever the FWHM spans at least $1.3$ bins; below that the
advantage shrinks without reversing
(Appendix~\ref{app:resolution}).

\emph{Experiment~S4} (Appendix~\ref{app:recovery}) repeats the comparison in
the time domain with the complete pipeline and exact inverse STFT on signals with known
components, adding VMD with the true $K$.  For stationary Lorentzian modes matching
$\alpha$, soft beats hard extraction in all $180$ paired trials.  For two tones and a chirp
the ranking depends on the SNR: at $0$\,dB the soft mask is better for all three, while
from $10$\,dB on the hard mask
recovers the two tones better ($0.078$ against $0.156$ for the $60$\,Hz tone at $20$\,dB),
because the prior FWHM of $31.6$\,Hz far exceeds the tones' width; for the chirp the soft
mask is still better at $10$\,dB and slightly worse at $20$\,dB.  VMD recovers the tones
at least as well as either, but the chirp best only at $0$\,dB.  The benefit thus depends on the prior
width matching the modes.

\emph{Experiment~S2} (Appendix~\ref{app:exp_e}) validates the FWHM
estimator of Corollary~\ref{cor:fwhm}: its accuracy is governed by the number of bins
across the FWHM, $n=2\beta/\Delta\omega$, so without noise $10\%$ is guaranteed only for
$n\geq22$; a smoothed spectrum is necessary but not sufficient; and the estimator
degrades once a neighbouring mode lies within about three FWHM.

\subsection{Experiment D: Real-World Signals}
\label{subsec:exp_f}

We evaluate on four real-world signal classes used
in~\cite{jia2025damd}, with identical preprocessing.
All use synchrosqueezed transforms: the SSQ-STFT or, for SSVEP, the SSQ-CWT on a
logarithmic axis.

\textbf{SSVEP EEG}~\cite{nakanishi2018}: the sample record distributed with the
TRCA method, a 40-target recording from a single subject whom
the distribution does not identify
(40 targets $\times$ 9 channels $\times$ 1250 samples $\times$ 6 blocks),
Oz electrode, averaged over the six blocks; the averaged responses to the
$10$\,Hz and $13$\,Hz targets (the third and sixth of the 40) are concatenated to give
5\,s of $10$\,Hz stimulation followed by 5\,s of $13$\,Hz; $f_s=250$\,Hz, $N=2500$,
SSQ-CWT.
\textbf{Gravitational wave}~\cite{g2net2021}: the first record of the G2Net data used
in~\cite{jia2025damd};
$f_s=220$\,Hz, $N=660$, SSQ-STFT.
\textbf{Bearing fault}~\cite{lessmeier2016}: bearing KA04 (type 6203, outer-race
damage) at 900\,rpm, $0.7$\,N\,m load torque and $1000$\,N radial force, first
measurement, vibration channel subsampled by
a factor of 256 (every 256th sample, without an anti-aliasing filter, as
in~\cite{jia2025damd}) from 64\,kHz; $f_s=250$\,Hz, $N=1000$, SSQ-STFT.
\textbf{Speech}~\cite{timit1993}: the TIMIT sample utterance distributed by the
LDC, first 32000 samples ($2$\,s at $16$\,kHz)
subsampled by a factor of 32 in the same way; $f_s=500$\,Hz, $N=1000$, SSQ-STFT.
Without low-pass filtering, energy above the new Nyquist rate aliases into the
band; these two records are what the pipeline receives.
SSQ-STFT uses a periodic Hann window of length
$n_\mathrm{fft}=f_s/2$; SSQ-CWT uses a Morlet wavelet with $64$
logarithmically spaced scales.
All four use $\mathrm{hop}=1$, percentile bandwidth estimation
with curvature adaptation, and $\alpha=2000$.
The implied prior FWHM $2/\sqrt{2\alpha}\approx0.032$ is $6.96$\,Hz
(gravitational wave), $7.91$\,Hz (SSVEP, bearing) and $15.8$\,Hz (speech); nothing was
tuned per dataset, and the $\alpha$-insensitivity of Experiment~C$'$ is not
established on these recordings.

\begin{figure*}[!t]
\centering
\includegraphics[width=\textwidth]{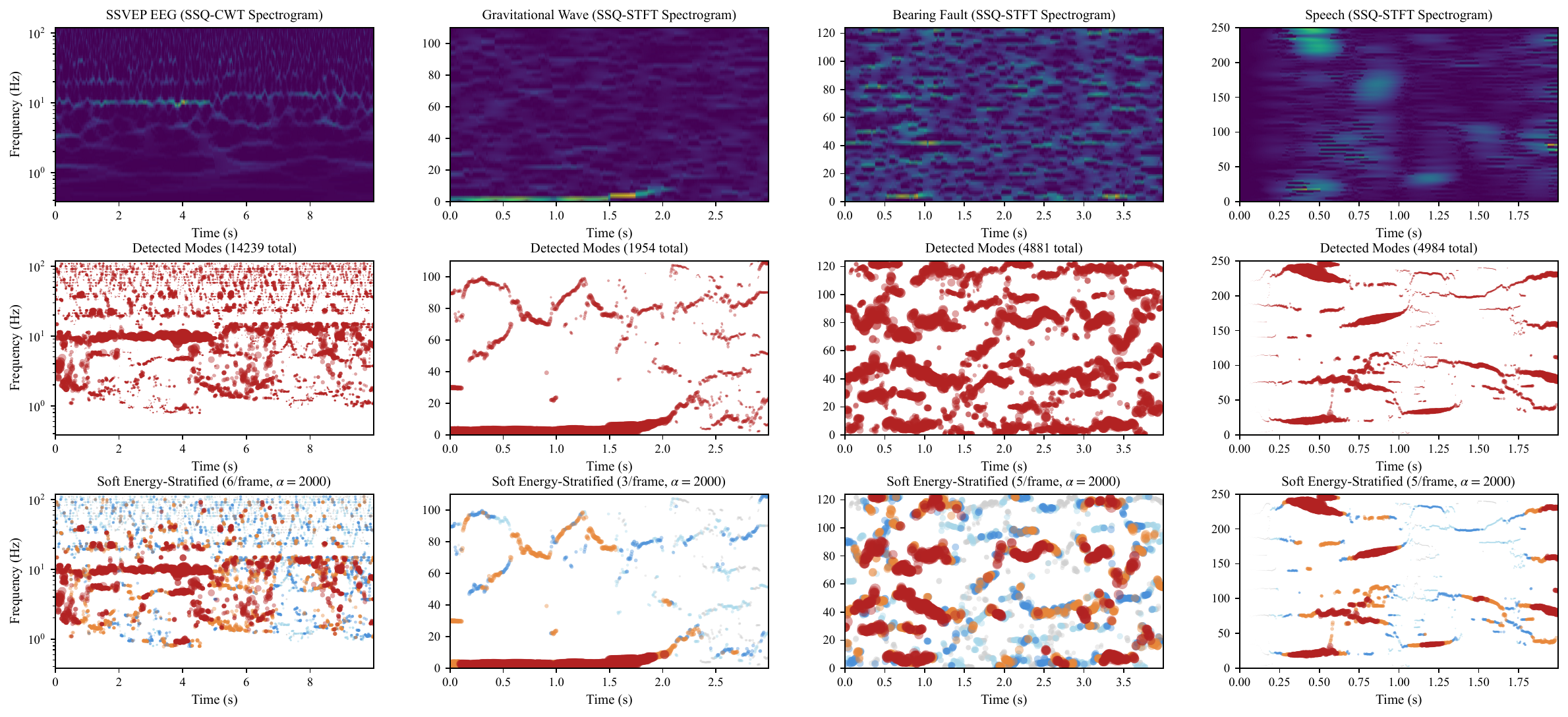}
\caption{Hard versus soft DAMD on four real-signal classes (Experiment~D).
Rows: spectrogram; hard detected modes (dot size $\propto$ energy); soft
extraction, centres coloured by energy quintile (energy-stratified), $\alpha=2000$.  All panels come from the
time--frequency map; no inverse transform is applied.
Settings: Section~\ref{subsec:exp_f}.  The per-frame counts in the third-row
titles are $\overline{K_t}$ of Table~\ref{tab:boundary}, rounded.}
\label{fig:exp_f}
\end{figure*}

Fig.~\ref{fig:exp_f} shows, per signal, the spectrogram, hard mode centres and
the soft result.
Table~\ref{tab:boundary} quantifies what the extraction rule affects.
\emph{Boundary}, the energy-weighted fraction of frame energy within one bin of a
Voronoi boundary, measures exposure only: both rules act on the same coefficients
there.  \emph{Allocation difference}, the quantity of interest, is
$\tfrac12\sum_k|s_k^\mathrm{soft}-s_k^\mathrm{hard}|P$ summed over the boundary bins
and divided by the frame energy: an input-power-weighted difference between the two
allocations, not energy transferred between reconstructed modes.  The masks act on
amplitude, $U_k=s_kX$, so modal power is $s_k^2P$: both rules preserve
$\sum_kU_k=X$, but modal powers sum to $P$ only where the allocation is
binary.  As the Lorentzian mask has full support, these figures are a lower bound that
grows with the neighbourhood width.
The allocation difference ranges from $1.03\%$ (gravitational wave) to $5.26\%$
(SSVEP) at a one-bin neighbourhood.
In the SSVEP column only $0.44$ of these $5.26$ percentage points come from bins
below $5$\,Hz; counting only coefficients above $5$\,Hz, where the $10$ and
$13$\,Hz responses lie, the one-bin figures are $13.60\%$ (boundary) and $5.91\%$
(allocation difference) (Appendix~\ref{app:realdata}).
Both grow monotonically with the radius $r$ in every column, the SSQ-STFT columns
keeping their order (bearing, speech, gravitational wave), so percentages must be read
with the width.  Columns are not directly
comparable: one bin is $2$\,Hz for SSQ-STFT but the SSQ-CWT axis is logarithmic
(widths in Appendix~H).  Neither quantity measures decomposition quality, which is assessed against ground
truth in Experiments~C and~C$'$; these recordings have none.

\begin{table}[t]
\centering
\caption{Boundary exposure and allocation difference (Experiment~D) for four
neighbourhood widths.  Grav.: gravitational wave; $\overline{K_t}$: mean
modes per frame.}
\label{tab:boundary}
\footnotesize
\setlength{\tabcolsep}{4pt}
\renewcommand{\arraystretch}{1.05}
\begin{tabular}{@{} l rrrr @{}}
\toprule
 & \textbf{SSVEP} & \textbf{Grav.} & \textbf{Bearing} & \textbf{Speech} \\
\midrule
 Transform            & SSQ-CWT & SSQ-STFT & SSQ-STFT & SSQ-STFT \\
$f_s$ (Hz)           & 250   & 220   & 250   & 500   \\
Duration (s)         & 10.00 & 3.00  & 4.00  & 2.00  \\
Frames               & 2500  & 660   & 1000  & 1000  \\
Bin width (Hz)       & 0.036 & 2.000 & 2.000 & 2.000 \\
$\overline{K_t}$     & 5.7   & 3.0   & 4.9   & 5.0   \\
Detections (total)   & 14239 & 1954  & 4881  & 4984  \\
\midrule
\multicolumn{5}{@{}l}{Boundary (\%), neighbourhood $\pm r$ bins} \\
\quad $r=1$          & 11.90 & 2.37  & 11.17 & 3.92  \\
\quad $r=2$          & 28.28 & 4.01  & 19.95 & 7.53  \\
\quad $r=3$          & 53.51 & 5.84  & 30.14 & 11.96 \\
\quad $r=5$          & 80.70 & 10.03 & 54.27 & 21.12 \\
\midrule
\multicolumn{5}{@{}l}{Allocation difference (\%), neighbourhood $\pm r$ bins} \\
\quad $r=1$          & 5.26  & 1.03  & 5.00  & 1.87  \\
\quad $r=2$          & 10.88 & 1.56  & 7.52  & 3.23  \\
\quad $r=3$          & 19.37 & 2.01  & 9.53  & 4.61  \\
\quad $r=5$          & 27.18 & 2.64  & 12.58 & 6.72  \\
\bottomrule
\end{tabular}

\vspace{2pt}
{\footnotesize SSQ-CWT bin width is the value at the low-frequency end of its
logarithmic axis; physical neighbourhood widths are in Appendix~H.}
\end{table}

\section{Discussion}
\label{sec:discussion}

DAMD~\cite{jia2025damd} addressed \emph{detection} through the clustering
equivalence; for \emph{extraction}, Theorems~\ref{thm:soft_opt}--\ref{thm:mmse} take the detected centres as given, from
any source, and derive the allocation optimal for the regularised
objective~\eqref{eq:soft_vmd}.  Hard Voronoi extraction optimises the linear
K-means functional, Lorentzian soft extraction the regularised quadratic VMD
functional, and the K-means and VMD objectives coincide \emph{only} at the simplex vertices (Theorem~\ref{thm:algebraic}).
The hard mask is not a limit of the Lorentzian family, whose $\alpha\to\infty$
endpoint is the VMD ADMM fixed point (Remark~\ref{rem:admm_fp}); the Lorentzian mask
applies to DAMD centres at no detection cost. Hard partition may suit impulsive
signals (untested here), soft partition overlapping modes whose width matches the
prior (Experiments~C$'$ and~S4); with detection reused, trying both is cheap.

In Experiments~C and~C$'$ hard extraction stays above the soft curve at exact
centres for all four tested spectral shapes, and for centre errors up to about one FWHM (Lorentzian), and
at every resolution tested; the soft mask loses at
high SNR at an inward centre error of two FWHM, or one FWHM
for non-Lorentzian shapes.  On Experiment~D the two allocations differ by $1.0\%$
to $5.0\%$ of the SSQ-STFT input power within one bin of a boundary.
Most of the advantage over hard extraction needs no regularisation: the nMSE of the
inverse-square allocation~\eqref{eq:vmd_wiener}, from the same centres, is within
$0.5\%$ of the mask's at every SNR of Experiment~C, and the gap grows as the modes
approach, to $0.8$--$3.7\%$ for modes within three FWHM (Experiment~C$'$).  Regularisation
therefore pays mainly for close modes.  Remark~\ref{prop:rate}
is a heuristic about the alternating iteration only, which Algorithm~\ref{alg:damd_soft}
does not run: Step~2 evaluates \eqref{eq:soft_opt} once at the detected centres and
relies on their accuracy, whose effect Experiment~C$'$ quantifies.

The parameter $\alpha$ comes from a known class value, the cheap FWHM estimator
(Corollary~\ref{cor:fwhm}; within $10\%$ once $n\geq22$ without noise, smoothing being
necessary but not sufficient under noise), or Type-II MLE,
most accurate on a noiseless model-matched spectrum (Appendix~C, Fig.~S2) but
iterative.

The Lorentzian model is approximate: for strongly non-stationary or impulsive
modes the MMSE equivalence of Theorem~\ref{thm:mmse} is lost, and the soft mask's
benefit depends on centre accuracy (Experiment~C$'$).
It also depends on the prior width: for line components at the default $\alpha$
it is worse than hard extraction from $10$\,dB on (Experiment~S4). Per-mode bandwidths are admitted by the theory
(Corollary~\ref{cor:weighted}) and estimated by Algorithm~\ref{alg:damd_soft} when
$\alpha$ is not supplied, but all experiments use a common $\alpha$, and the amplitude
weights $\sigma_k^2$ of \eqref{eq:weighted_opt} are not estimated.  The framework is
single-channel; multi-channel aggregation is an independent direction.

\section{Conclusion}
\label{sec:conclusion}

The VMD--K-means equivalence rests on the idempotency of binary weights; on the
simplex interior the objectives diverge, and the
regularised soft VMD problem has a closed-form optimum, unique on the support of the
spectrum: the Lorentzian weights.
They \emph{bridge} VMD and DAMD: every VMD ADMM fixed point is $\propto
|\omega-\omega_k|^{-2}$ and, with reconstruction enforced, is the $\alpha\to\infty$
endpoint of the Lorentzian family, which finite $\alpha$ regularises; and they equal the MMSE estimator of each mode in the low-noise
limit under a Lorentzian spectral prior (at finite noise the exact MMSE estimator
adds a shrinkage factor).
Hard Voronoi and Lorentzian soft extraction are optima of two different objectives
over the same simplex, both needing only the detected centres: the choice is a modelling decision.
In the experiments most of the gain over hard extraction is already given by the
inverse-square allocation~\eqref{eq:vmd_wiener} from the same centres; a finite $\alpha$
adds a few per cent for close modes.

Two limitations remain.  Mode labels are not associated across frames, so
globally labelled waveform modes are out of scope.  And the MMSE statement is
coefficient-domain: reconstruction transfers to the waveform for
any invertible transform, but per-bin optimality does not, because a masked coefficient
array need not lie in the range of the analysis operator
(Remark~\ref{rem:timedomain}), and the synchrosqueezed transforms used here are
nonlinear.

\section*{Data and Code Availability}
Code reproducing Experiments~A--C$'$ and S1--S5, and applying the measurement of
Experiment~D to synthetic signals, is provided with the published article.  The four
recordings of Experiment~D are not included; they are available from their distributors
\cite{nakanishi2018,g2net2021,lessmeier2016,timit1993}, and the record, channel and segment
used are given in Section~\ref{subsec:exp_f}.
The code is also publicly available at\\
{\def\UrlBreaks{\do\/\do\-}\url{https://github.com/plustar/A-Closed-Form-Soft-Spectral-Partition-for-Adaptive-Mode-Decomposition}}.

\clearpage
\noindent
The appendices below are the supplementary material of the published article: the proofs
(Appendix~\ref{app:proofs}), Experiments S1--S5
(Appendices~\ref{app:exp_b}--\ref{app:runtime}), and supporting detail for
Remark~VII.1, Table~IV and Experiment~C$'$ (Appendices~\ref{app:stft}--\ref{app:resolution}).
Their equations, figures and table are numbered with the prefix S.

\renewcommand{\theequation}{S\arabic{equation}}
\setcounter{equation}{0}
\renewcommand{\thefigure}{S\arabic{figure}}
\setcounter{figure}{0}
\renewcommand{\thetable}{S\arabic{table}}
\setcounter{table}{0}

\appendices

\section{Proofs}
\label{app:proofs}

\subsection*{Theorem~III.1 (Algebraic Identity)}
\begin{proof}
($\Leftarrow$)\; If $s_k(\omega)\in\{0,1\}$, then $s_k^2(\omega) = s_k(\omega)$ pointwise, so the integrands of $J_\mathrm{VMD}$ and $J_\mathrm{Kmeans}$ coincide at every $\omega$, and the integral identity holds for any signal $X$ and any centres $\{\omega_k\}$.

($\Rightarrow$)\; Fix the centres $\{\omega_k\}$ and the allocation functions
$s_k$, which by hypothesis do not depend on the signal, and suppose
$J_\mathrm{VMD}=J_\mathrm{Kmeans}$ for every signal $X$.  (The quantifier over $X$
cannot be dropped: for a single $X$ the identity constrains $s_k$ only where
$|X|^2>0$.)  Let $\omega_0>0$ be any point with $\omega_k\neq\omega_0$ for
all $k$.  For $0<\varepsilon<2\omega_0$, so that the spectrum below is supported in
$(0,\infty)$ and is that of an admissible analytic signal, choose $|X_\varepsilon(\omega)|^2 = \varepsilon^{-1} \mathbf{1}_{[\omega_0 - \varepsilon/2,\, \omega_0 + \varepsilon/2]}(\omega)$, a unit-mass spectrum concentrated at $\omega_0$.  Then
\begin{align}
0 \;=\; J_\mathrm{VMD}^{(\varepsilon)} - J_\mathrm{Kmeans}^{(\varepsilon)}
   \;=\;& \frac{1}{\varepsilon}\int_{\omega_0 - \varepsilon/2}^{\omega_0 + \varepsilon/2}
   \sum_{k=1}^K |\omega - \omega_k|^2\notag\\
   &\times \bigl[s_k^2(\omega) - s_k(\omega)\bigr]\, d\omega.
\label{eq:S_fixed_signal}
\end{align}
Taking $\varepsilon \to 0$, the Lebesgue differentiation theorem gives
\begin{equation}
\sum_{k=1}^K |\omega_0 - \omega_k|^2 \bigl[s_k^2(\omega_0) - s_k(\omega_0)\bigr] = 0 \quad \text{for a.e.\ } \omega_0>0.
\label{eq:S_pointwise}
\end{equation}
Since $s_k(\omega_0) \in [0,1]$ implies $s_k^2(\omega_0) \leq s_k(\omega_0)$, each term in the sum is non-positive, and the sum vanishes only if each term individually vanishes.  For $\omega_0$ chosen so that $|\omega_0 - \omega_k|^2 > 0$ for all $k$ (which holds outside the discrete set $\{\omega_k\}$), this requires $s_k^2(\omega_0) = s_k(\omega_0)$, i.e., $s_k(\omega_0) \in \{0,1\}$.  Since the exceptional set has Lebesgue measure zero, $s_k \in \{0,1\}$ almost everywhere
on $[0,\infty)$, the domain of $s_k$.
\end{proof}

\subsection*{Corollary~III.2 (Strict Inequality for Soft Allocations)}
\begin{proof}
On any such set, $s_k^2(\omega)<s_k(\omega)$, while
$|\omega-\omega_k|^2P(\omega)>0$.  Hence the corresponding integrand
of $J_\mathrm{VMD}-J_\mathrm{Kmeans}$ is strictly negative on a set of
positive measure, and all other terms are non-positive.
\end{proof}

\subsection*{Theorem~IV.3 (Closed-form optimum of regularised soft VMD)}
\begin{proof}
Fix $\omega$ with $P(\omega)>0$.  The per-frequency
problem~(11) reduces to
\begin{equation}
  \min_{s\in\Delta^K} \sum_{k=1}^K c_k s_k^2,
  \quad c_k = \Bigl(|\omega-\omega_k|^2+\tfrac{1}{2\alpha}\Bigr)P(\omega)>0,
\label{eq:local_qp}
\end{equation}
where $\Delta^K=\{s:s_k\geq 0,\,\sum_k s_k=1\}$ is the standard
$(K-1)$-simplex.  The objective $\sum_k c_k s_k^2$ is strictly convex
on $\Delta^K$ (positive definite diagonal Hessian, $c_k>0$) and
$\Delta^K$ is compact and convex, so the minimum is attained uniquely.

The minimiser is interior: if $s_k=0$ and $s_j>0$ at a feasible point, moving
mass $\varepsilon\in(0,s_j]$ from $s_j$ to $s_k$ changes the objective by
$-2c_js_j\varepsilon+(c_j+c_k)\varepsilon^2$, which is negative for small
$\varepsilon$.  Hence every $s_k>0$ at the minimiser, the non-negativity constraints
are inactive, and the KKT conditions reduce to
\begin{equation}
  2c_k s_k = \lambda \quad \forall k,
\end{equation}
giving $s_k = \lambda/(2c_k)$.  Substituting into $\sum_k s_k=1$:
\begin{equation}
  \lambda = \frac{2}{\sum_{j=1}^K c_j^{-1}},
\end{equation}
so $s_k^* = c_k^{-1}/\sum_j c_j^{-1}$, which after substituting
$c_k = (|\omega-\omega_k|^2+1/(2\alpha))P(\omega)$ and cancelling the
common factor $P(\omega)$ yields (12).  At frequencies
where $P(\omega)=0$ the integrand vanishes for any $s_k$ and the
extension by~(12) contributes zero to $\tilde J_\alpha$.
\end{proof}

\subsection*{Corollary~IV.7 (Weighted Generalisation)}
\begin{proof}
Fix $\omega$ with $P(\omega)>0$.  The per-frequency cost of (14) is
$\tilde c_k=(d_k^2+\beta_k^2)P(\omega)/(\sigma_k^2\beta_k^2)>0$, with
$d_k=|\omega-\omega_k|$ and $\beta_k^2=1/(2\alpha_k)$.  By the same KKT argument as
Theorem~IV.3, $s_k^*=\tilde c_k^{-1}/\sum_j\tilde c_j^{-1}$.  Here
\begin{align*}
  \tilde c_k^{-1}
  &= \frac{\sigma_k^2\beta_k^2}{(d_k^2+\beta_k^2)P(\omega)}
   = \frac{\sigma_k^2}{\bigl(1+d_k^2/\beta_k^2\bigr)P(\omega)}\\
  &= \frac{\sigma_k^2}{\bigl(1+2\alpha_kd_k^2\bigr)P(\omega)},
\end{align*}
so $P(\omega)$ cancels between numerator and denominator and (15) follows.  Where
$P(\omega)=0$ the integrand vanishes for any feasible $s_k$, and (15) is adopted by
convention, as in Theorem~IV.3.  Note that
dividing by $\sigma_k^2$ alone would leave
$\tilde c_k^{-1}=2\alpha_k\sigma_k^2/[(1+2\alpha_kd_k^2)P]$, i.e.\ an extra factor
$\alpha_k$ that (15) does not contain and that cancels only when the bandwidths are
equal; this is the error corrected in Remark~IV.8.
\end{proof}

\subsection*{Corollary~IV.9 (Limits of the Lorentzian family)}
\begin{proof}
Let $d_j:=|\omega-\omega_j|>0$ for all $j$.  As $\alpha\to\infty$,
$[1+2\alpha d_j^2]^{-1}=(2\alpha d_j^2)^{-1}(1+o(1))$, so the common
$1/(2\alpha)$ prefactor cancels in numerator and denominator:
\begin{equation}
  s_k^*(\omega)
  = \frac{d_k^{-2}(1+o(1))}{\sum_{j=1}^K d_j^{-2}(1+o(1))}
  \;\longrightarrow\;
  \frac{d_k^{-2}}{\sum_{j=1}^K d_j^{-2}}.
\label{eq:S_limit_ratio}
\end{equation}
Expanding one order further, $[1+2\alpha d_j^2]^{-1}=(2\alpha
d_j^2)^{-1}[1-(2\alpha d_j^2)^{-1}+O(\alpha^{-2})]$, so the difference between
$s_k^*(\omega)$ and its limit is $O(1/\alpha)$: it decays with $\alpha$ rather than
vanishing beyond a finite threshold.
At $\omega=\omega_k$ the formula (12) gives
$[1+0]^{-1}=1$ in the numerator, while $[1+2\alpha d_j^2]^{-1}\to 0$
for $j\neq k$ \emph{provided} $d_j>0$, i.e.\ provided the other centres are distinct
from $\omega_k$; hence $s_k^*(\omega_k)\to 1$.  If $m$ centres coincide at
$\omega$, each of those $m$ numerators equals one and the remaining ones tend to zero,
so each of the coincident modes receives $1/m$.  The $\alpha\to 0$ limit is
immediate since, for each fixed $\omega$, $[1+2\alpha d_j^2]^{-1}\to 1$ for every
$j$, giving
equal mass $1/K$ to every mode.
\end{proof}

\subsection*{Theorem~V.2 (MMSE Equivalence)}
\begin{proof}
The MMSE estimator for complex Gaussian observations is the Wiener--Hopf solution \cite{wiener1949extrapolation,kay1993fundamentals}:
\begin{equation*}
  \hat{U}_k^\mathrm{MMSE}(\omega) = \frac{S_{U_k^\mathrm{true}, X}(\omega)}{S_X(\omega)}\, X(\omega).
\end{equation*}
By independence of the modes (Definition~V.1), the cross-spectrum is $S_{U_k^\mathrm{true}, X} = S_k$, giving (20).  In the low-noise limit, $S_X \to \sum_j S_j$, and substituting the Lorentzian PSDs~(19):
\begin{align*}
  \frac{S_k(\omega)}{\sum_j S_j(\omega)}
  &= \frac{\sigma_k^2 [1 + (\omega - \omega_k)^2/\beta_k^2]^{-1}}{\sum_j \sigma_j^2 [1 + (\omega - \omega_j)^2/\beta_j^2]^{-1}}\\
  &= \frac{\sigma_k^2 [1 + 2\alpha_k (\omega - \omega_k)^2]^{-1}}{\sum_j \sigma_j^2 [1 + 2\alpha_j (\omega - \omega_j)^2]^{-1}} = s_k^*(\omega),
\end{align*}
where $\beta_k^2 = 1/(2\alpha_k)$ and $s_k^*$ is the weighted soft partition~(15).  For $\sigma_k^2 = \sigma^2$, $\alpha_k = \alpha$, the $\sigma^2$ factors cancel and the result reduces to~(12).
\end{proof}

\subsection*{Proposition~VI.1 (Soft M-step)}
\begin{proof}
Setting $\partial \tilde J_\alpha/\partial\omega_k=0$ with
$s_k$ held fixed:
\begin{align}
  \frac{\partial}{\partial\omega_k}
  \int \!\Bigl(|\omega-\omega_k|^2+\tfrac{1}{2\alpha}\Bigr)
  s_k^2 P\,d\omega
  &= -2\!\int (\omega-\omega_k) s_k^2 P\,d\omega \notag\\
  &= 0,
\end{align}
which gives (28) directly.  The $1/(2\alpha)$ term
contributes a zero derivative and therefore plays no role in the
centre update.
\end{proof}

\subsection*{Proposition~VI.2 (Monotone decrease of the objective)}
\begin{proof}
The E-step minimises $\tilde J_\alpha$ over $\{s_k\}\in\Delta^K$
for fixed $\{\omega_k^{(n)}\}$, with a global minimiser given
by Theorem~IV.3; so
$\tilde J_\alpha^{(n+1/2)}\leq \tilde J_\alpha^{(n)}$.  The M-step
minimises $\tilde J_\alpha$ over $\{\omega_k\}$ for fixed $s^{(n+1)}$,
with minimiser given by the centroid formula~(28)
(Proposition~VI.1); so
$\tilde J_\alpha^{(n+1)}\leq \tilde J_\alpha^{(n+1/2)}$.  Together,
$\tilde J_\alpha^{(n+1)}\leq \tilde J_\alpha^{(n)}$, and the
non-negative non-increasing sequence converges.
\end{proof}

\subsection*{Remark~VI.7 (Heuristic sensitivity estimate)}
\begin{proof}[Derivation sketch]
The composed E--M map $\Phi:\omega_k^{(n)}\mapsto\omega_k^{(n+1)}$ is
differentiable near the fixed point.  Writing
$a_j=[1+2\alpha(\omega-\omega_j)^2]^{-1}$, $S=\sum_\ell a_\ell$ and
$s_k^*=a_k/S$, the quotient rule gives the E-step sensitivity
\begin{equation}
  \frac{\partial s_k^*}{\partial\omega_j}
  = s_k^*\bigl(\delta_{jk}-s_j^*\bigr)\,
    \frac{4\alpha(\omega-\omega_j)}{1+2\alpha(\omega-\omega_j)^2},
\label{eq:estep_sensitivity}
\end{equation}
hence, since $s_k^*(1-s_k^*)\le s_k^*$ and $s_k^*s_j^*\le s_j^*$,
\begin{equation}
  \Bigl|\frac{\partial s_k^*}{\partial\omega_j}\Bigr|
  \;\le\; s_j^*(\omega)\,g_j(\omega),
\label{eq:estep_bound}
\end{equation}
where $g_j(\omega)=4\alpha|\omega-\omega_j|/[1+2\alpha(\omega-\omega_j)^2]\le\sqrt{2\alpha}$.
A bound of the form
$C\alpha|\omega-\omega_j|[1+2\alpha(\omega-\omega_j)^2]^{-2}$ with $C$ depending only
on $K$ does not hold, because it requires $\sum_\ell a_\ell$ to be bounded
below, which fails far from all centres (the ratio of the two sides grows with
$\alpha$).  Differentiating the M-step~(28),
\begin{equation}
  \frac{\partial\omega_k^{(n+1)}}{\partial\omega_j^{(n)}}
  = \frac{\int(\omega-\omega_k^{(n+1)})\,2s_k^*\,
          (\partial s_k^*/\partial\omega_j)\,P\,d\omega}
         {\int[s_k^*]^2P\,d\omega}.
\label{eq:mstep_jacobian}
\end{equation}
The denominator must be bounded away from zero (condition~(ii) of Remark~VI.7).  In
the numerator, \eqref{eq:estep_bound} makes the integrand small wherever
$s_j^*$ or $s_k^*$ is small; near well-separated centres this happens within a
neighbourhood of width $O(1/\sqrt{\alpha})$ of each centre (for $j=k$ near $\omega_k$,
through the factor $1-s_k^*$ of \eqref{eq:estep_sensitivity}), so the contribution of
power concentrated near those centres may be small.  Far from all centres, the
normalised weights need not be concentrated on one mode; hence the products of weights
in \eqref{eq:mstep_jacobian} need not be small.  Control of that region depends on the
spectral mass $P$ there (condition~(iii)).  The argument is therefore a
heuristic under conditions~(i)--(iii), not a bound: it does not establish
monotonicity in $\Delta\omega_{\min}$, contraction or a convergence rate, which would
require a sign-aware spectral-radius computation that is not attempted.
\end{proof}

\section{Experiment S1: Implementation Consistency}
\label{app:exp_b}

\subsection*{Setup and result}
\label{subsec:exp_b}

\textit{Purpose}: Confirm that the closed-form Lorentzian weights
(Theorem~IV.3) and the normalised Lorentzian mask~(6), implemented
independently, agree to floating-point precision.  By Remark~IV.5 the
coincidence holds by construction, so this is a consistency check between two
implementations, not a validation.

\textit{Setup}: Two-tone signal ($f_1=80$\,Hz, $f_2=200$\,Hz,
$\alpha=2000$).  DAMD detects the two mode centres; the closed-form
$s_k^*(\omega)$ and the mask $w_k(\omega)$ of~(6) are
evaluated at the same centres and compared pointwise.

\textit{Result}: $\|s_k^*-w_k\|_\infty$ is at the floating-point floor, which
confirms only that the two implementations evaluate the same analytic expression
identically.  It is a unit test rather than evidence for any claim about the ADMM
solver; the relationship to ADMM is tested in Experiment~A.

\begin{figure}[t]
\centering
\includegraphics[width=0.6\textwidth]{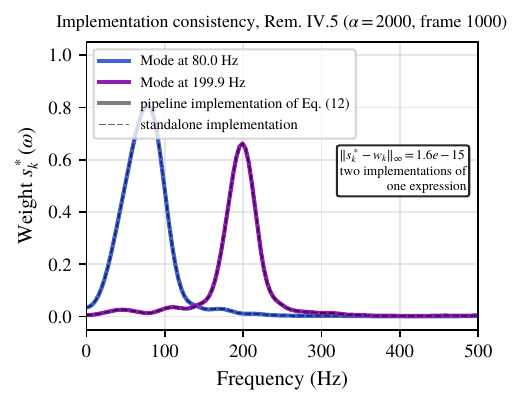}
\caption{Lorentzian soft-partition weights $s_k^*(\omega)$ from the pipeline
implementation and the normalised Lorentzian mask $w_k(\omega)$ of~(6) from a standalone
implementation, for two modes, $\alpha=2000$ (Experiment~S1).  The maximum difference
$\|s_k^*-w_k\|_\infty$, printed in the panel, is at the
floating-point floor, consistent with the by-construction
identity in Remark~IV.5.}
\label{fig:exp_b}
\end{figure}

\section{Experiment S2: The FWHM Estimator}
\label{app:exp_e}

\subsection*{Setup and result}
\label{subsec:exp_e}

\textit{Purpose}: establish the regime in which Corollary~V.5 is accurate,
which is governed by the number of bins across the FWHM rather than by $\alpha$
itself.

\textit{Setup}: a synthetic Lorentzian mode, $\alpha_\mathrm{true}$ swept over four
decades, results reported against $n=2\beta/\Delta\omega$.  Since the FWHM is an integer
bin count $m$ with $|m-n|<1$, the error obeys
$|\delta\alpha/\alpha|\le(1-1/n)^{-2}-1\approx2/n$ in the worst case over the position
of the centre within a bin and, to first order, $4f(1-f)/n\le1/n$ on average ($f$ the
fractional part of $n$); both curves are plotted.  Panel (a) reports the error at a
single centre ($\omega_0=0.30$) and as the worst of $20$ sub-bin offsets.  Three
estimators are compared: FWHM (Corollary~V.5), the power-weighted second moment, and
the Type-II MLE of (27), which treats the centre, mode power and noise floor as known
and optimises $\log\alpha$ alone by a bounded Brent solver on the fixed interval
$\alpha\in[1,10^7]$ (tolerance $10^{-4}$, no starting value or random
initialisation).  Equation (27) is written for a common $\alpha$; with one mode the common
$\alpha$ is the per-mode $\alpha_k$, and with several modes the gradient with
respect to $\alpha_k$ keeps only the $k$-th term of $\partial S_X/\partial\alpha$, with
$\sigma_k^2$ and $\alpha_k$.  Panel (b) repeats the measurement on noisy periodograms
averaged over $L\in\{1,4,16,64\}$ realisations.

\textit{Result}: the FWHM estimator follows the $1/n$ envelope within the
worst-case bound.  At the single centre position the largest error with $n\geq10$ is
$8.7\%$; over the $20$ offsets, errors above $10\%$ occur up to $n\approx12$ and reach
$17.7\%$.  A continuous scan over $n$ and $400$ offsets
never exceeds the bound, agrees with $4f(1-f)/n$
to within $0.013$ for $n\geq10$, and shows the worst case exceeding $10\%$ for some $n$
up to about $20$, just below integers; $10\%$ is guaranteed only for $n\geq22$.  The
second moment is unusable, as expected for a Lorentzian, which has none.  Under noise a
single periodogram is far too rough for a half-power crossing, and averaging is
necessary but not sufficient at $20$\,dB: after $L=64$ averages the median error is
still $49\%$ for $n=25.6$ and $31\%$ for $n=11.4$, and $5\%$ only for the narrowest
mode.  A smoothed spectrum is a precondition of Corollary~V.5, and the per-frame
magnitudes used in Algorithm~1 are single periodograms.

\textit{Overlap and centre error}.  With a
second equal-amplitude Lorentzian $s$ FWHM away and the band taken to the Voronoi
midpoint, the error is unchanged for $s\gtrsim3$ and then grows: for $n=11.4$ and the
neighbour above the mode it is $8.3\%$ isolated, $9.0\%$ at $s=3$, $22.4\%$ at $s=2$ and
$41.7\%$ at $s=1.5$; for $n=5.1$, $4.9\%$ isolated and $27.2\%$ at $s=1.5$.  With the
neighbour below, the grid position shifts these by a bin ($n=11.4$: $9.0\%$ at $s=2$,
$33.1\%$ at $s=1.5$; $n=5.1$: $27.2\%$ and $46.5\%$).  The estimator thus inherits the
resolution limit of its band and should not be relied on below about three FWHM.  A
centre error of up to $1$ FWHM changes the count by at most one bin ($8.3\%$ or $9.0\%$
at $s=3$, $n=11.4$).  A $2$-FWHM common displacement fails in two ways: towards the
neighbour the band admits its peak and the error is about $75\%$; away from it the
Voronoi midpoint truncates the mode's own half-power region and the error is $424\%$.
The band must isolate the mode; given that, the estimate is insensitive to where inside
the band the centre is thought to be.

\begin{figure*}[t]
\centering
\includegraphics[width=0.8\textwidth]{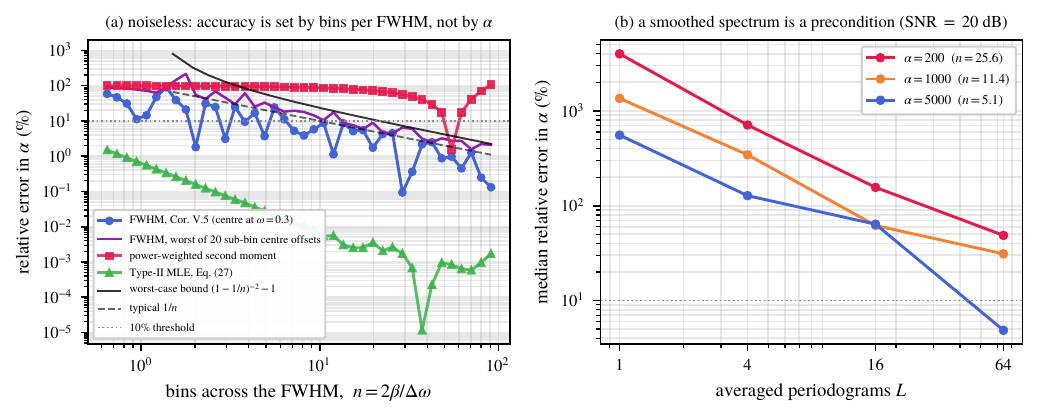}
\caption{Accuracy of the FWHM estimator (Experiment~S2,
Corollary~V.5).  (a) Noiseless case, plotted against the number of bins
across the FWHM, $n=2\beta/\Delta\omega$, rather than against $\alpha$: since the FWHM
is measured as an integer bin count, the average error over the position of the centre
is at most about $1/n$ (dashed) and the
worst case over that position is $(1-1/n)^{-2}-1$ (solid
black).  At one centre position the $10\%$ threshold is met from $n\approx10$; over $20$
sub-bin offsets it is exceeded up to $n\approx12$ at the sampled $n$ (up to $n\approx20$
on a continuous scan), and guaranteed only for $n\geq22$.  The
power-weighted second moment is unusable throughout because a Lorentzian has no finite
second moment; the Type-II MLE is far more accurate on a noiseless spectrum matching
its model, at the cost of a numerical solve.  (b) With noise: a single periodogram is
too rough for a half-power crossing, and spectral averaging improves but does not by
itself restore the estimator at $20$\,dB, so a smoothed spectrum is a precondition of
Corollary~V.5, one that has to be supplied by averaging or smoothing and that the
single-frame magnitudes used in Algorithm~1 do not by themselves provide.}
\label{fig:exp_e}
\end{figure*}

\section{Experiment S3: Numerical Verification of Theorem~IV.3}
\label{app:closed_form}

The closed form of Theorem~IV.3 is compared against a generic sequential quadratic
programme on the simplex, so that the check does not re-evaluate the expression under
test.  Over $400$ random configurations with $K\in[2,6]$ and $\alpha\in[10,10^5]$ the
two agree to a median of $6\times10^{-9}$ and a maximum of $3\times10^{-7}$; at the
closed-form point the stationarity conditions $2c_ks_k^*=\mathrm{const}$ hold to
$9\times10^{-19}$; and none of $20\,000$ random feasible perturbations attains a lower
objective value.

\begin{figure*}[t]
\centering
\includegraphics[width=0.8\textwidth]{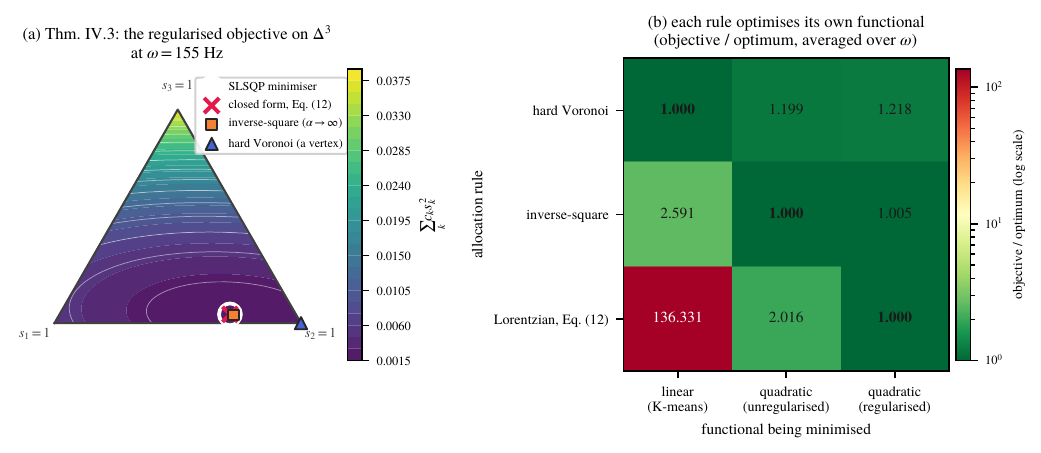}
\caption{Theorem~IV.3 and the three functionals.
(a) The regularised objective (11) over the two-simplex at one
frequency, for $K=3$.  The closed form (12) (cross) sits on the
minimiser found by a generic sequential quadratic programme (circle); the
inverse-square allocation and the hard Voronoi mask, which is a vertex, are marked for
comparison.  Over $400$ random configurations with $K\in[2,6]$ and
$\alpha\in[10,10^5]$ the two agree to a median of $6\times10^{-9}$.
(b) Each allocation rule evaluated under each functional, normalised by that
functional's own optimum and averaged over frequency: every rule attains one under its
own functional and strictly more under the others.  The hard mask optimises the linear
K-means functional, the inverse-square allocation the unregularised quadratic one, and
(12) the regularised quadratic one, which is
Theorem~III.1 and Corollary~III.2 read as a table.}
\label{fig:supp_closed_form}
\end{figure*}

\section{Experiment S4: Time-Domain Mode Recovery with Known Components}
\label{app:recovery}
\textit{Purpose}: Experiments C and C$'$ compare the extraction rules in a
frequency-domain Gaussian model.  This experiment runs the complete pipeline (STFT,
density-based detection, extraction and inverse STFT) on time-domain signals whose
components are known, and scores the recovered waveforms against them.

\textit{Signals}: $f_s=1000$\,Hz, $N=1024$, $x=u_1+u_2+u_3+$ white Gaussian noise, with
SNR (total signal power over noise power) of $0$, $10$ and $20$\,dB and $20$ trials per
SNR (seed $42$).  Two classes share the spacing of $u_1$ and $u_2$.  \emph{Lorentzian
modes}: stationary Gaussian processes with Lorentzian spectra centred at $60$, $110$ and
$250$\,Hz and half-width $\beta=f_s/\sqrt{2\alpha}=15.8$\,Hz, the model of
Definition~V.1 with a width matching $\alpha=2000$, each realisation scaled to RMS
$(1,0.8,0.6)/\sqrt2$.  \emph{Line components}: $\cos(2\pi60t+\phi_1)$,
$0.8\cos(2\pi110t+\phi_2)$ and the chirp $0.6\cos(2\pi(200t+50t^2)+\phi_3)$ from $200$ to
$302$\,Hz, with random phases.

\textit{Methods}: the pipeline of Experiment~S5 (STFT, $n_\mathrm{fft}=128$, hop $1$), run
once per trial; the same detected centres are used by the soft mask~(12) with
$\alpha=2000$ and by the hard mask~(7), and the masked STFT is inverted exactly.  The
hard bands of the pipeline coincide with the Voronoi mask~(7) of the detected centres in
every one of the $4.0\times10^6$ (bin, frame) cells checked.  VMD is run on the whole
signal with the true $K=3$, its penalty matched to $\alpha=2000$ in the convention of
Section~II, and its default $\tau=10^{-5}$, initialisation and tolerance.
No parameter was tuned for this experiment.

\textit{Association}: the pipeline returns $K_t$ unlabelled modes per frame, on average
$9.0$ (line components) and $9.5$ (Lorentzian modes) for three true components.  Each
detected mode is assigned to the component whose true instantaneous frequency at that
frame is nearest, and the masks assigned to a component are summed.  This uses the true
instantaneous frequencies and is identical for the soft and the hard mask; VMD modes are
matched to components by centre frequency.  Both methods thus receive oracle
information (the pipeline the association, VMD the number of modes), and the absolute
errors should be read with that in mind.  The comparison of soft with hard extraction,
which differ only in the mask, is not affected by it.

\textit{Metric}: $\|\hat u_c-u_c\|/\|u_c\|$ per component, mean $\pm$ standard error over
trials, and the number of trials in which the soft error is below the hard error on the
same detection.  In every trial the soft and the hard estimates sum to $x$ to machine
precision, as a partition of unity must.

\begin{table}[h]
\centering
\caption{Experiment S4: relative waveform error (mean $\pm$ standard error, $20$
trials).  Last column: trials in which soft $<$ hard on the same detection.}
\label{tab:recovery}
\footnotesize
\setlength{\tabcolsep}{1.5pt}
\begin{tabular}{@{}r@{\hspace{6pt}}l c c c r@{}}
\toprule
SNR & component & soft & hard & VMD ($K=3$) & soft$<$hard \\
\midrule
\multicolumn{6}{@{}l}{\textit{Lorentzian modes matching $\alpha$}} \\
0 & 60 Hz & 0.722\,$\pm$\,0.007 & 0.781\,$\pm$\,0.007 & 0.703\,$\pm$\,0.007 & 20/20 \\
0 & 110 Hz & 0.888\,$\pm$\,0.011 & 1.001\,$\pm$\,0.013 & 0.833\,$\pm$\,0.020 & 20/20 \\
0 & 250 Hz & 1.878\,$\pm$\,0.013 & 1.935\,$\pm$\,0.013 & 0.966\,$\pm$\,0.035 & 20/20 \\
10 & 60 Hz & 0.523\,$\pm$\,0.004 & 0.553\,$\pm$\,0.004 & 0.625\,$\pm$\,0.006 & 20/20 \\
10 & 110 Hz & 0.654\,$\pm$\,0.005 & 0.706\,$\pm$\,0.006 & 0.693\,$\pm$\,0.007 & 20/20 \\
10 & 250 Hz & 0.752\,$\pm$\,0.004 & 0.772\,$\pm$\,0.004 & 0.643\,$\pm$\,0.007 & 20/20 \\
20 & 60 Hz & 0.504\,$\pm$\,0.006 & 0.529\,$\pm$\,0.006 & 0.631\,$\pm$\,0.005 & 20/20 \\
20 & 110 Hz & 0.631\,$\pm$\,0.007 & 0.672\,$\pm$\,0.007 & 0.693\,$\pm$\,0.006 & 20/20 \\
20 & 250 Hz & 0.521\,$\pm$\,0.004 & 0.534\,$\pm$\,0.005 & 0.620\,$\pm$\,0.005 & 20/20 \\
\midrule
\multicolumn{6}{@{}l}{\textit{Line components (tones and chirp)}} \\
0 & tone 60 Hz & 0.530\,$\pm$\,0.006 & 0.578\,$\pm$\,0.006 & 0.319\,$\pm$\,0.006 & 20/20 \\
0 & tone 110 Hz & 0.669\,$\pm$\,0.010 & 0.771\,$\pm$\,0.011 & 0.559\,$\pm$\,0.068 & 20/20 \\
0 & chirp & 1.793\,$\pm$\,0.007 & 1.847\,$\pm$\,0.007 & 0.965\,$\pm$\,0.025 & 20/20 \\
10 & tone 60 Hz & 0.218\,$\pm$\,0.002 & 0.185\,$\pm$\,0.002 & 0.123\,$\pm$\,0.003 & 0/20 \\
10 & tone 110 Hz & 0.275\,$\pm$\,0.003 & 0.247\,$\pm$\,0.003 & 0.142\,$\pm$\,0.003 & 0/20 \\
10 & chirp & 0.582\,$\pm$\,0.002 & 0.595\,$\pm$\,0.002 & 0.709\,$\pm$\,0.002 & 20/20 \\
20 & tone 60 Hz & 0.156\,$\pm$\,0.001 & 0.078\,$\pm$\,0.003 & 0.076\,$\pm$\,0.004 & 0/20 \\
20 & tone 110 Hz & 0.201\,$\pm$\,0.002 & 0.098\,$\pm$\,0.003 & 0.083\,$\pm$\,0.006 & 0/20 \\
20 & chirp & 0.203\,$\pm$\,0.002 & 0.193\,$\pm$\,0.002 & 0.685\,$\pm$\,0.001 & 0/20 \\
\bottomrule
\end{tabular}
\end{table}

\textit{Result}: For Lorentzian modes matching $\alpha$, the soft mask has the lower
error in all $180$ paired trials.  Against VMD it is better at $20$\,dB for all three
modes and at $10$\,dB for two of them, and worse at $0$\,dB.  For line components the
ranking depends on the SNR: the soft mask is better at $0$\,dB, but from $10$\,dB on the
hard mask recovers both tones better, because a tone spans one or two bins while the
soft mask at $\alpha=2000$ has an FWHM of $31.6$\,Hz and leaks part of each tone into the
mode $50$\,Hz away, as in Experiment~C$'$, where a mask much wider than the modes falls
behind hard extraction.  Corollary~V.5 cannot repair this: one or two bins across the
FWHM is far below the $n\approx10$ at which the errors of Appendix~C fall to about
$10\%$, and the $n\geq22$ that guarantees $10\%$ without noise.  VMD, given the true $K$, has the
lowest tone errors at every SNR (for the $60$\,Hz tone at $20$\,dB within one standard
error of hard extraction), but its fixed centres cannot follow the chirp, on which it
is behind both masks from $10$\,dB on; at $0$\,dB it is ahead there as well, for the
reason given in the next paragraph.

At $0$\,dB the highest-frequency component has errors above one for both masks.  About two
thirds of the spurious detections ($64\%$ and $63\%$ at $0$\,dB) lie above the highest
true component and are assigned to it, and since the masks form a partition of unity every out-of-band noise
bin ends up in some component.  VMD's modes do not sum to $x$ at the iteration cap with
$\tau=10^{-5}$: they leave a residual of $0.62$ (line components) and $0.68$ (Lorentzian
modes) of $\|x\|$ at $0$\,dB, correlated with the noise at $0.83$ and $0.76$, which is
why VMD is ahead at that SNR.

\section{Experiment S5: Computational Cost}
\label{app:runtime}
Its conclusions are summarised in Section~VII-B.

\textit{Setup}: a three-component non-stationary signal (broadband noise would
leave the detection stage nothing to find), $f_s=1000$\,Hz, $n_\mathrm{fft}=128$,
$\mathrm{hop}=1$, $N\in\{64,\ldots,2048\}$; medians of nine repetitions after one
untimed warm-up, with interquartile ranges.  STVMD [5] uses $K=10$, dynamic mode
tracking and a penalty of $50$ on a frequency axis $(n_\mathrm{fft}/F)\,\omega$,
$F=\lfloor n_\mathrm{fft}/2\rfloor+1$, which in the convention of Section~II is
$\alpha=50\,(n_\mathrm{fft}/F)^2/2\approx97$.
The soft stage is timed by forming the $(K_t\times F)$ weight matrix of (12) for
every frame and applying it to that frame's slice, the $O(T\cdot F\cdot K_t)$ work
Table~II counts.  No reconstruction is timed for DAMD, whose per-frame modes have no
waveform counterpart without an association of labels across frames (Section~X);
the STVMD timing includes its inverse STFT.  Both methods run on the CPU of an
Intel Core Ultra 9 288V laptop (eight cores, 32\,GB) under Windows~11,
single-threaded, with Python 3.14, NumPy 2.5, SciPy 1.18 and Numba 0.67, which
compiles the mean-shift loop; no GPU is used.

\begin{figure*}[t]
\centering
\includegraphics[width=0.8\textwidth]{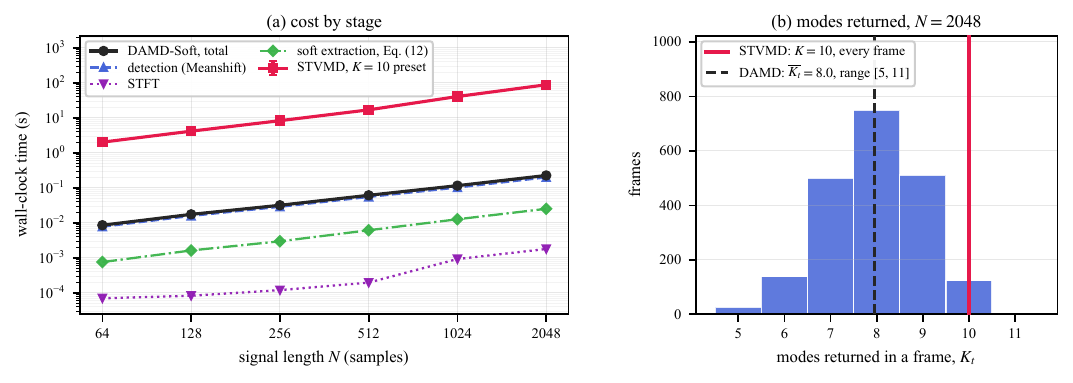}
\caption{Computational cost (Experiment~S5); medians of nine repetitions after a
warm-up, bars are interquartile ranges.  (a) Cost by stage.  (b) Modes returned at the
largest length, against the fixed $K$ given to STVMD.}
\label{fig:exp_g}
\end{figure*}

\textit{Result}: the end-to-end ratio rises from $236\times$ at $N=64$ to
$390\times$ at $N=2048$ (Fig.~\ref{fig:exp_g}(a)).  Both methods scale about linearly in
$T=N$, so the gap is the constant factor $K\cdot N_\mathrm{A}$ of the ADMM loop, and
the ratio should be read as a range rather than a single speed-up.  Soft extraction
accounts for $9$--$11\%$ of the DAMD pipeline; detection dominates.  STVMD is given
$K=10$ in every frame, whereas DAMD returns $K_t\in[5,11]$, mean $8.0$
(Fig.~\ref{fig:exp_g}(b)), more than the three synthesised components, so the ratio is a
pipeline comparison between methods with different mode-selection rules.

\textit{Output quality}: the outputs cannot be scored mode by mode, since STVMD
returns $K=10$ tracked modes and the present algorithm $K_t$ unassociated modes per
frame.  Both must sum back to the signal: at $N=512$ the soft allocation does so to
$3.7\times10^{-16}$, the round-trip error of the transform, and STVMD to
$9.1\times10^{-3}$ at its iteration cap of $500$ ($499$ ADMM updates).  STVMD runs
with its default dual step $\tau=10^{-5}$, for which the iterate at the cap is still close
to the shrunk inverse-square form of Section~II (its $\tau=0$ case), so this residual reflects $\tau$ as well as the
iteration count; its $\alpha$ is not matched to the soft mask's, and how its cost
varies with $\alpha$ was not examined.  The accuracy of the allocation itself is
measured against known modes in Experiments~C and~C$'$.

\section{The STFT Pair and the Range of the Masked Array}
\label{app:stft}
This appendix supports Remark~VII.1.  \textit{Analysis and synthesis}: the signal is
reflect-padded so that frame $t$ is centred on sample $t\cdot\mathrm{hop}$; each frame of $n_\mathrm{fft}$
samples is multiplied by a periodic Hann window and transformed by a real FFT without
further scaling, with the phase referenced to the frame centre, so only the $F=\lfloor n_\mathrm{fft}/2\rfloor+1$ non-negative
frequencies are stored and masked.  Synthesis applies the inverse real FFT to each
masked frame, which restores Hermitian symmetry and returns real-valued (not analytic)
modes, and combines the frames by windowed overlap-add divided by the overlap-added
squared window.

\textit{Measurements} ($n_\mathrm{fft}=256$, unit hop, $\alpha=2000$).  The round trip is exact to a relative error of
$\sim6\times10^{-16}$.  Transferring per-bin MMSE optimality to the waveform needs the
analysis operator to be an isometry up to a constant \emph{and} the masked array to
remain in its range.  The first holds to within $0.7\%$ ($\|\mathcal{T}x\|^2/\|x\|^2$
across test signals); the second does not: the analyse--synthesise--analyse residual of $s_k^*\mathcal{T}x$ is
$1.2$--$2.1\times10^{-3}$ for a three-tone signal and $1.7$--$3.2\times10^{-2}$ for
white noise over the three modes, so per-bin optimality does not transfer exactly; the
per-bin independence of Definition~V.1 is in any case an idealisation here.

\section{Experiment D: The SSVEP Column and the Neighbourhood Width}
\label{app:realdata}
This appendix gives the detail behind Table~IV in Section~VIII-D.

\textit{Metadata.}  With unit hop there is one frame per sample, so the frame count
is $N$.  The synchrosqueezed STFT keeps the DFT grid $kf_s/n_\mathrm{fft}$, and with
$n_\mathrm{fft}=f_s/2$ one bin is $2.000$\,Hz in all three SSQ-STFT columns, including the
bearing column with its odd $n_\mathrm{fft}=125$.  The SSQ-CWT axis
is logarithmic: consecutive bins differ by a factor of $1.0955$, and $0.036$\,Hz is the
spacing at the low-frequency end.

\textit{The SSVEP column.}  Of its $5.26\%$ allocation difference at $r=1$, $0.44$
percentage points come from bins below $5$\,Hz.  With $\alpha$ fixed in normalised
frequency the mask FWHM is about $7.9$\,Hz, and $37\%$ of adjacent centre pairs on the
logarithmic axis lie within one FWHM ($35\%$ of these below $5$\,Hz), where the soft
weights are shared among several modes.  Counting only coefficients above $5$\,Hz
(numerator and denominator alike), where the $10$ and $13$\,Hz responses lie, the
one-bin figures are $13.60\%$ (boundary) and $5.91\%$ (allocation difference).

\textit{Neighbourhood width.}  Both quantities grow monotonically with $r$ in every
column, and the SSQ-STFT columns keep the same order at every radius (bearing, speech,
gravitational wave).  In those columns the allocation difference as a share of the boundary energy falls
from $44$--$48\%$ at $r=1$ to $23$--$32\%$ at $r=5$, so the difference between the
rules is concentrated nearest the boundary; in the SSVEP column it falls from $44\%$ to $34\%$,
and the neighbourhoods cover, on average over frames, $22\%$ of the bins at $r=1$ and
$61\%$ at $r=5$.  For the SSQ-STFT
columns a $\pm r$-bin neighbourhood is $(2r+1)$ bins, i.e.\ $6$, $10$, $14$ and $22$\,Hz
for $r=1,2,3,5$; on the logarithmic SSQ-CWT axis the same neighbourhoods span frequency
ratios of $1.31$, $1.58$, $1.89$ and $2.73$ wherever the boundary lies.

\section{Experiment C$'$: Frequency Resolution}
\label{app:resolution}
The per-bin draws of Experiments~C and~C$'$ do not model window leakage, which is
what a coarse resolution does in practice.  Here the two modes are time series (complex
first-order autoregressions with unit peak and, near the peak, the half-width
$\beta=0.01$ of the mask, $\alpha=5000$, centred at $0.12$ and $0.30$) analysed by a
Hann-windowed STFT with $n_\mathrm{fft}$ from $16$ to $1024$ (hop $n_\mathrm{fft}/4$),
i.e.\ $0.32$ to $20.5$ bins across the FWHM; by linearity, the STFT of each mode is the
exact ground truth, and both masks use the exact centres.  Over $40$ records of $8192$ samples, the ratio
$\mathrm{nMSE}_\mathrm{soft}/\mathrm{nMSE}_\mathrm{hard}$ is $0.79$ at $40$\,dB, $0.74$
at $20$\,dB and $0.71$ at $0$\,dB for every $n_\mathrm{fft}\geq64$ (at least $1.3$ bins
across the FWHM), and changes by less than $0.01$ between $n_\mathrm{fft}=64$ and $1024$.
Below that the advantage shrinks at high SNR, to $0.81$ at $0.64$ bins and $0.92$ at
$0.32$ bins across the FWHM ($40$\,dB), without reversing.  Standard errors of the ratio
are below $0.002$.

\bibliographystyle{IEEEtran}
\bibliography{main}

\begin{thebibliography}{10}
\providecommand{\url}[1]{#1}
\csname url@samestyle\endcsname
\providecommand{\newblock}{\relax}
\providecommand{\bibinfo}[2]{#2}
\providecommand{\BIBentrySTDinterwordspacing}{\spaceskip=0pt\relax}
\providecommand{\BIBentryALTinterwordstretchfactor}{4}
\providecommand{\BIBentryALTinterwordspacing}{\spaceskip=\fontdimen2\font plus
\BIBentryALTinterwordstretchfactor\fontdimen3\font minus
  \fontdimen4\font\relax}
\providecommand{\BIBforeignlanguage}[2]{{%
\expandafter\ifx\csname l@#1\endcsname\relax
\typeout{** WARNING: IEEEtran.bst: No hyphenation pattern has been}%
\typeout{** loaded for the language `#1'. Using the pattern for}%
\typeout{** the default language instead.}%
\else
\language=\csname l@#1\endcsname
\fi
#2}}
\providecommand{\BIBdecl}{\relax}
\BIBdecl

\bibitem{oppenheim1999discrete}
A.~V. Oppenheim and R.~W. Schafer, \emph{Discrete-Time Signal Processing},
  2nd~ed.\hskip 1em plus 0.5em minus 0.4em\relax Upper Saddle River, NJ, USA:
  Prentice Hall, 1999.

\bibitem{dragomiretskiy2014vmd}
K.~Dragomiretskiy and D.~Zosso, ``Variational mode decomposition,'' \emph{IEEE
  Transactions on Signal Processing}, vol.~62, no.~3, pp. 531--544, 2014.

\bibitem{wiener1949extrapolation}
N.~Wiener, \emph{Extrapolation, Interpolation, and Smoothing of Stationary Time
  Series}.\hskip 1em plus 0.5em minus 0.4em\relax MIT Press, 1949.

\bibitem{boyd2011distributed}
S.~Boyd, N.~Parikh, E.~Chu, B.~Peleato, and J.~Eckstein, ``Distributed
  optimization and statistical learning via the alternating direction method of
  multipliers,'' \emph{Foundations and Trends in Machine Learning}, vol.~3,
  no.~1, pp. 1--122, 2011.

\bibitem{jia_short-time_2025}
H.~Jia, P.~Cao, T.~Liang, C.~F. Caiafa, Z.~Sun, Y.~Kushihashi, A.~Grau,
  Y.~Bolea, F.~Duan, and J.~Sol{\'e}-Casals, ``Short-time variational mode
  decomposition,'' \emph{Signal Processing}, vol. 238, p. 110203, 2026.

\bibitem{nazari_successive_2020}
M.~Nazari and S.~M. Sakhaei, ``Successive variational mode decomposition,''
  \emph{Signal Processing}, vol. 174, p. 107610, 2020.

\bibitem{liu_successive_2022}
S.~Liu and K.~Yu, ``Successive multivariate variational mode decomposition,''
  \emph{Multidimensional Systems and Signal Processing}, vol.~33, no.~3, pp.
  917--943, 2022.

\bibitem{lian_adaptive_2018}
J.~Lian, Z.~Liu, H.~Wang, and X.~Dong, ``Adaptive variational mode
  decomposition method for signal processing based on mode characteristic,''
  \emph{Mechanical Systems and Signal Processing}, vol. 107, pp. 53--77, Jul.
  2018.

\bibitem{xu_adaptive_2023}
C.~Xu, J.~Yang, T.~Zhang, K.~Li, and K.~Zhang, ``Adaptive parameter selection
  variational mode decomposition based on a novel hybrid entropy,''
  \emph{Measurement}, vol. 217, p. 113110, 2023.

\bibitem{wang_adaptive_2022}
J.~Wang, C.~Zhan, S.~Li, Q.~Zhao, J.~Liu, and Z.~Xie, ``Adaptive variational
  mode decomposition based on {A}rchimedes optimization algorithm,''
  \emph{Measurement}, vol. 191, p. 110798, 2022.

\bibitem{yan_adaptive_2024}
M.~Yan, L.~Chen, W.~Hu, Y.~Chen, and X.~Zhou, ``An adaptive variational mode
  decomposition algorithm based on improved particle swarm optimization,'' in
  \emph{2024 IEEE 6th International Conference on Power, Intelligent Computing
  and Systems (ICPICS)}, 2024, pp. 670--675.

\bibitem{gu_time-frequency_2025}
Z.~Gu, Z.~Wang, Y.~Shen, C.~Huang, Y.~Jiang, and P.~Li, ``Time-frequency
  information-based variational mode decomposition,'' \emph{Measurement Science
  and Technology}, vol.~36, p. 036122, 2025.

\bibitem{jia2025damd}
H.~Jia, C.~Yang, C.~F. Caiafa, Z.~Sun, F.~Duan, and J.~Sol{\'e}-Casals,
  ``Density-based adaptive mode decomposition,'' \emph{IEEE Transactions on
  Signal Processing}, vol.~74, pp. 1368--1382, 2026.

\bibitem{comaniciu2002mean}
D.~Comaniciu and P.~Meer, ``Mean shift: A robust approach toward feature space
  analysis,'' \emph{IEEE Transactions on Pattern Analysis and Machine
  Intelligence}, vol.~24, no.~5, pp. 603--619, 2002.

\bibitem{cheng1995mean}
Y.~Cheng, ``Mean shift, mode seeking, and clustering,'' \emph{IEEE Transactions
  on Pattern Analysis and Machine Intelligence}, vol.~17, no.~8, pp. 790--799,
  1995.

\bibitem{fukunaga1975estimation}
K.~Fukunaga and L.~D. Hostetler, ``The estimation of the gradient of a density
  function, with applications in pattern recognition,'' \emph{IEEE Transactions
  on Information Theory}, vol.~21, no.~1, pp. 32--40, 1975.

\bibitem{srinivasan2006binary}
S.~Srinivasan, N.~Roman, and D.~Wang, ``Binary and ratio time-frequency masks
  for robust speech recognition,'' \emph{Speech Communication}, vol.~48,
  no.~11, pp. 1486--1501, 2006.

\bibitem{wang2005ideal}
D.~Wang, ``On ideal binary mask as the computational goal of auditory scene
  analysis,'' in \emph{Speech Separation by Humans and Machines}, P.~Divenyi, Ed.\hskip 1em
  plus 0.5em minus 0.4em\relax Boston, MA, USA: Springer, 2005, pp. 181--197.

\bibitem{kay1993fundamentals}
S.~M. Kay, \emph{Fundamentals of Statistical Signal Processing, Volume~{I}:
  Estimation Theory}.\hskip 1em plus 0.5em minus 0.4em\relax Englewood Cliffs,
  NJ, USA: Prentice Hall, 1993.

\bibitem{papoulis2002probability}
A.~Papoulis and S.~U. Pillai, \emph{Probability, Random Variables, and
  Stochastic Processes}, 4th~ed.\hskip 1em plus 0.5em minus 0.4em\relax
  McGraw-Hill, 2002.

\bibitem{dempster1977maximum}
A.~P. Dempster, N.~M. Laird, and D.~B. Rubin, ``Maximum likelihood from
  incomplete data via the {EM} algorithm,'' \emph{Journal of the Royal
  Statistical Society: Series B}, vol.~39, no.~1, pp. 1--38, 1977.

\bibitem{nakanishi2018}
M.~Nakanishi, Y.~Wang, X.~Chen, Y.-T. Wang, X.~Gao, and T.-P. Jung, ``Enhancing
  detection of {SSVEPs} for a high-speed brain speller using task-related
  component analysis,'' \emph{IEEE Transactions on Biomedical Engineering},
  vol.~65, no.~1, pp. 104--112, 2018.

\bibitem{g2net2021}
\BIBentryALTinterwordspacing
{European Gravitational Observatory}, ``{G2Net} gravitational wave detection,''
  2021. [Online]. Available:
  \url{https://www.kaggle.com/competitions/g2net-gravitational-wave-detection}
\BIBentrySTDinterwordspacing

\bibitem{lessmeier2016}
C.~Lessmeier, J.~K. Kimotho, D.~Zimmer, and W.~Sextro, ``Condition monitoring
  of bearing damage in electromechanical drive systems by using motor current
  signals of electric motors: A benchmark data set for data-driven
  classification,'' in \emph{Proceedings of the European Conference of the
  Prognostics and Health Management Society}, vol.~3, no.~1, 2016, pp. 1--17.

\bibitem{timit1993}
J.~S. Garofolo, L.~F. Lamel, W.~M. Fisher, J.~G. Fiscus, and D.~S. Pallett,
  ``{TIMIT} acoustic-phonetic continuous speech corpus,'' {LDC93S1}, Linguistic
  Data Consortium, Philadelphia, PA, USA, 1993. [Online]. Available:
  \url{https://catalog.ldc.upenn.edu/LDC93S1}

\end{thebibliography}

\end{document}